\documentclass{amsart}
\usepackage{amssymb, verbatim, color, graphicx, booktabs, natbib}
\usepackage{amsmath}
\usepackage{subfigure} 

\newcommand{\alphab}{\boldsymbol \alpha}
\newcommand{\zerob}{\mathbf 0}
\newcommand{\Ub}{\mathbf U}
\newcommand{\yb}{\mathbf y}
\newcommand{\Bb}{\mathbf B}
\newcommand{\Db}{\mathbf D}
\newcommand{\Qb}{\mathbf Q}
\newcommand{\Rb}{\mathbf R}
\newcommand{\Ib}{\mathbf I}
\newcommand{\Vb}{\mathbf V}
\newcommand{\Lb}{\mathbf L}
\newcommand{\B}{\boldsymbol}

\usepackage[margin=1in,a4paper]{geometry}
\bibpunct{(}{)}{;}{a}{}{,}

\usepackage{natbib}

\makeatletter
\def\@biblabel#1{\@ifnotempty{#1}{}}
\makeatother

\begin{document}

\title[Efficient flexible regression for groundwater monitoring]{Efficient and automatic methods for flexible regression on spatiotemporal data, with applications to groundwater monitoring}

\author{A.\ W.\ Bowman}
\address[A.\ W.\ Bowman, L.\ Evers,\ D.\ Molinari]{School of Mathematics and Statistics, University of Glasgow, 15 University Gardens, Glasgow, G12 8QQ}
\author{L.\ Evers}
\author{D.\ Molinari}
\author{W.\ R.\ Jones}
\address[W.\ R.\ Jones, M.\ J.\ Spence]{Shell Global Solutions, UK}
\author{M.\ J.\ Spence}

\begin{abstract}
Fitting statistical models to spatiotemporal data requires finding the right balance between imposing smoothness and following the data. In the context of p-splines, we propose a Bayesian framework for choosing the smoothing parameter which allows the construction of fully-automatic data-driven methods for fitting flexible models to spatiotemporal data.  A computationally efficient implementation, exploiting the sparsity of the arising design and penalty matrices, is proposed.  The findings are illustrated using a simulation and two examples, all concerned with the modelling of contaminants in groundwater, which suggest that the proposed strategy is more stable that competing strategies based on the use of criteria such as GCV and AIC.
\end{abstract}


\maketitle

\section{Introduction}
\label{sec:introduction}

Spatiotemporal data have become ubiquitous.  In some settings this has been driven by the development of affordable technology for data collection where spatially located networks of sensors collect data over time.  In environmental monitoring multiple sensors are routinely used to gather data over time, in air, water or land settings.  Brain imaging using {\sc eeg} (electro-encephalography) or {\sc meg} (magneto-encephalography) is another example where around $200$ sensors each record brain signals at very high time resolution, generating large volumes of data.  In many scientific contexts, measurements are increasingly made automatically, leading to high resolution data with a strong degree of regularity, while on other occasions visits to sites of interest by trained personnel may be required, leading to sparser and more irregular data patterns.  The problem discussed in the present paper deals with measurements of groundwater collected from wells and sent for subsequent lab analysis.  The practicalities and cost of this inevitably lead to irregularity in time and also in space, even when operating within a fixed set of sampling locations determined by the well positions.

Models for the analysis and interpretation of spatiotemporal data have developed rapidly to match the demands of the data now available and the underlying questions.  Sometimes prediction is the aim while on other occasions interest can be directed at assessing the mean levels of the measurement and evidence for change over time.  \citet{banerjee-2004-book}, \citet{finkenstadt-2007-book} and \citet{cressie-2011-book} provide excellent entry points to the large literature on spatiotemporal modelling, with the last book very helpfully giving coverage of modern hierarchical and dynamic methods in both breadth and depth.  These models are usually implemented in a Bayesian setting.  In the wider literature, a unifying theme is the expectation that the spatial and temporal patterns exhibited will not follow simple parametric forms, so that models which can express flexible, but generally smooth, shapes are required.  One approach is to apply flexible forms of regression, described for example by \citet{wood-2006-book}, in the spatiotemporal setting.  \citet{bowman-2009-applstat} take this approach to the modelling of sulphur dioxide over Europe throughout the 1990's.  P-splines, described by \citet{eilers-1996-statsci}, and more general regression splines, offer a very interesting approach through the use of relatively low-dimensional sets of basis functions and \citet{lee-2011-statmod} apply this to the spatiotemporal modelling of ozone over Europe.  The formulation of p-splines offers an interpretation in terms of mixed effects and \citet{ruppert-2003-book} showed the wide range of settings to which these models can be applied when the random effect interpretation is appropriate.  A fully Bayesian p-splines model was introduced by \citet{lang-2004-jcgs}, with inference carried out by {\sc mcmc}.  \citet{fahrmeir-2004-statsinica} adopted a model of this type in the specific setting of spatiotemporal data, with an empirical Bayes approach which returns again to a mixed-model representation.  \citet{brezger-2006-csda} provided a wider range of models and efficient updating schemes while \citet{brezger-2008-statmod} discussed simultaneous probability statements for Bayesian p-spline models, again in the context of {\sc mcmc} implemetation.  More recently, \citet{wood-2011-jrssb} explored the {\sc reml} approach in detail and developed a fast implementation in a generalised linear modelling framework.  

The context of the application discussed in this paper is the monitoring of contamination in groundwater. It is clearly important to assess water quality and its associated risks to human health and the wider environment, and in particular to detect sudden increases in contaminant concentration due to possible releases.  The contaminants in the groundwater are measured using water samples collected from wells and sent for subsequent lab analysis.  The practicalities and cost of this inevitably lead to irregularity in time and also in space, even when operating within a fixed set of sampling locations determined by the well positions. The data collection and assessment activity is generally undertaken by staff who have science or engineering background, but may not have had advanced training in statistical methods. However it is impractical that results should always be referred back to others for statistical analysis and so there is a practical need for statistical tools that can be implemented easily and robustly as a routine part of the work of those environmental professionals. The analysis therefore needs to be fully automatic and to be fast to carry out, but also to produce results which are reliable, informative, and aid robust project decision-making.

The aim of the present paper is to address these issues.  In order to allow the construction of flexible models over space and time, p-splines are used because of their ability to provide compact representations and to express smoothness control in simple forms, as described in Section~\ref{sec:psplines}.  A fully Bayesian spatiotemporal model is introduced in Section~\ref{sec:bayesian}, using conjugate priors to avoid the need for {\sc mcmc} implementation.  In particular, the issue of selecting the degree of smoothness in the model is also addressed in order to produce a fully automatic procedure.  A focus will be on issues of `ballooning', where predictions can be high in areas where there is no data, and this is identified and addressed by appropriate choices of the number of basis functions and the type of smoothness penalty used.  The need for speed is addressed through matrix decompositions which enable the parameter which controls smoothness to be separated out from the computationally intensive parts of the calculation, along similar lines to those used by \citet{ruppert-2003-book}, but also exploiting the sparsity of the design matrices associated with the spline basis.

\section{Spatiotemporal smoothing by p-splines}
\label{sec:psplines}

The p-spline approach to smoothing has become widely used because of its simplicity and its `low rank' representation of the function of interest.  In the simplest case where responses $y_i$ and covariate values $x_i$ are observed for a sample $i = 1, \ldots, n$, the model $y_i = m(x_i) + \varepsilon_i$ describes a flexible underlying relationship through the nonparametric regression function $m$ whose form is unspecified beyond an assumption of smoothness.  A basis approach assumes this function can be expressed through a linear combination $\sum_{j=1}^p \alpha_j \phi_j(x)$, where the functions $\phi_j(x)$ are usually taken to be b-splines (usually of order $3$) because of their efficient construction from polynomial pieces.  By modifying the values of the coefficients $\alpha_j$ a huge range of smooth functions can be created by weighting the local p-spline basis functions which are centred at a grid of values along the x-axis.

A simple method of extending this to the spatiotemporal setting, where data $y_i$ are indexed over space $(s_{1i}, s_{2i})$ and time $t_i$, for $i = 1, \ldots, n$, is to express the regression function as $m(s_1, s_2, t) = \sum_j \sum_k \sum_l \alpha_{jkl} \phi_j(s_1) \phi_k(s_2) \phi_l(t)$.  This uses a basis set which is simply the product of all triples of the marginal basis functions over $s_1$, $s_2$ and $t$.  This can be conveniently expressed in vector-matrix form as $\mathbf{Y} = \mathbf{B} \boldsymbol \alpha + \boldsymbol \varepsilon$, where $\boldsymbol Y$ and $\boldsymbol \varepsilon$ are vectors of response data and error terms, $\boldsymbol \alpha$ is the vector of parameters $\alpha_{jkl}$ and the design matrix $\mathbf{B}$ consists of the basis functions (columns) evaluated at each data point (rows).

Although this model can be fitted by simple least squares, \citet{eilers-1996-statsci} proposed to use a dense set of basis functions in conjunction with a penalty term to control the degree of smoothness in the estimate.  Specifically, the parameter estimate is chosen to be the value of $\alphab$ which minimises
\begin{equation}\label{eqn:objective}
 \left\|\yb - \Bb \alphab \right\|^2 + \lambda \left\|\Db \alphab\right\|^2 ,
\end{equation}
where the matrix $\Db$ computes successive differences across the sequence of $\alpha_{jkl}$'s in each of the three covariate dimensions.  Second-order differences are often used.  Large values of the smoothing parameter $\lambda$ thereby induce smoothness in the values of $\alphab$ and hence in the estimated function $m$.  The solution for the basis coefficients is easily seen to be $\hat{\alphab} = ( \Bb^T\Bb + \lambda \Db^T\Db)^{-1} \Bb^T \yb$.  The trace of the matrix $\Bb (\Bb^T\Bb + \lambda \Db^T\Db )^{-1} \Bb^T$ which creates the fitted values from the data vector $\yb$ is defined as the `effective degrees of freedom' by analogy with standard linear models.  This gives a more intuitive scale on which the smoothness of the estimate can be expressed.  The details of these methods are described by \citet{eilers-1996-statsci}, \citet{ruppert-2003-book}, \citet{wood-2006-book} and many other authors.  In particular, \citet{lee-2011-statmod} discuss p-splines in the spatiotemporal setting.

In all forms of flexible or nonparametric regression, the choice of the degree of smoothness for the estimator is a crucial one and many authors have addressed this issue.  Widely used approaches include cross-validation ({\sc cv}) which, using spatiotemporal notation, chooses $\lambda$ to minimise $\sum_{i=1}^n \{y_i - \hat{m}_{-i}(s_{1i}, s_{2i}, t_i)\}^2$, where the subscript $-i$ indicates that observation $i$ is not included in the construction of the estimate.  Generalised cross-validation ({\sc gcv}) is a popular variation.  Akaike's Information Criterion ({\sc aic}) and its variation {\sc aic}c, described by \citet{hurvich-1989-aicc}, are also widely used as a means of balancing the goodness-of-fit of the model against its complexity.  The general thinking is that {\sc aic}c is less affected by the under-smoothing to which {\sc aic} and cross-validation are sometimes prone. Other commonly used criteria are {\sc gcv} and {\sc bic}. The details of all these methods are discussed by many authors, with \citet{wood-2006-book} a good starting point.

\begin{figure}
\centerline{
   \begin{tabular}{ll}
\subfigure[{\sc aic}c (all wells)]{\includegraphics[width = 0.45\textwidth]{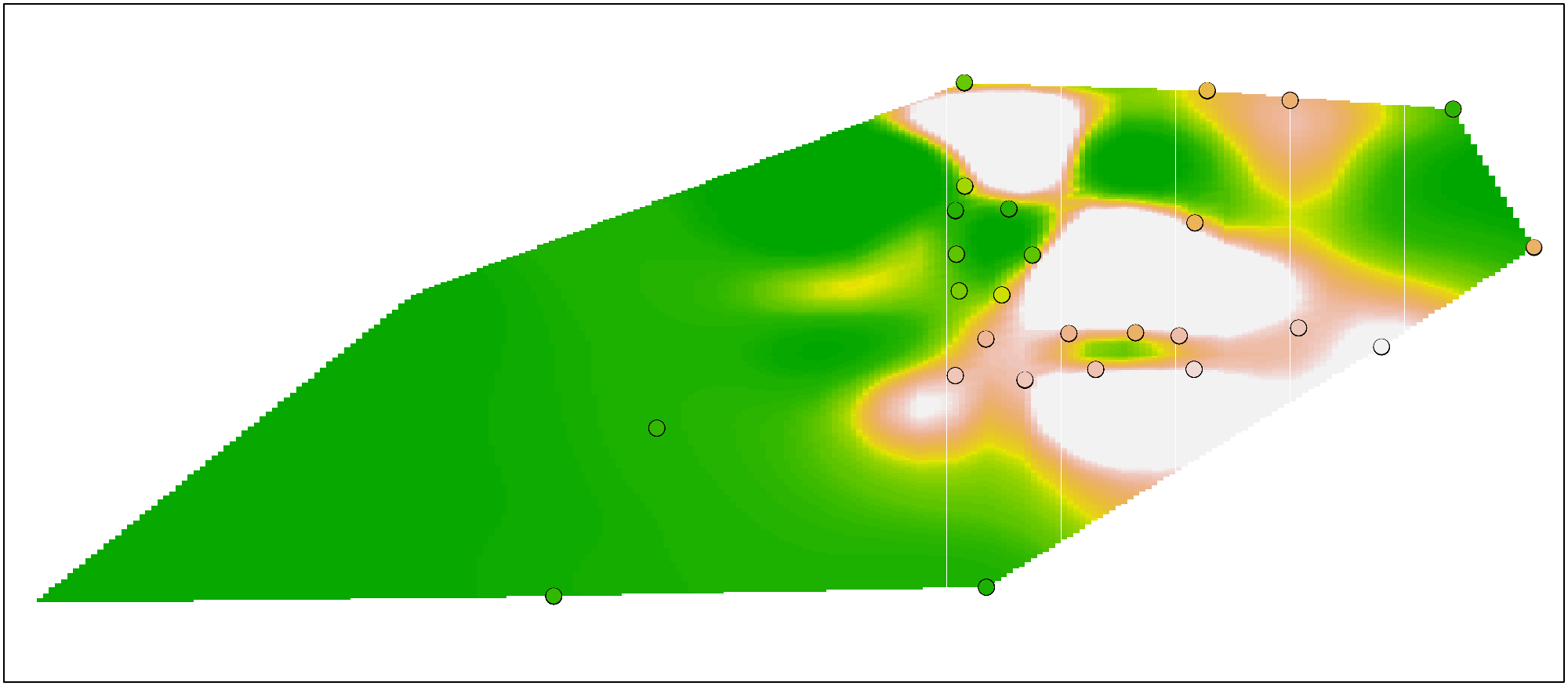}}
\subfigure[{\sc aic}c (four wells removed)]{\includegraphics[width = 0.45\textwidth]{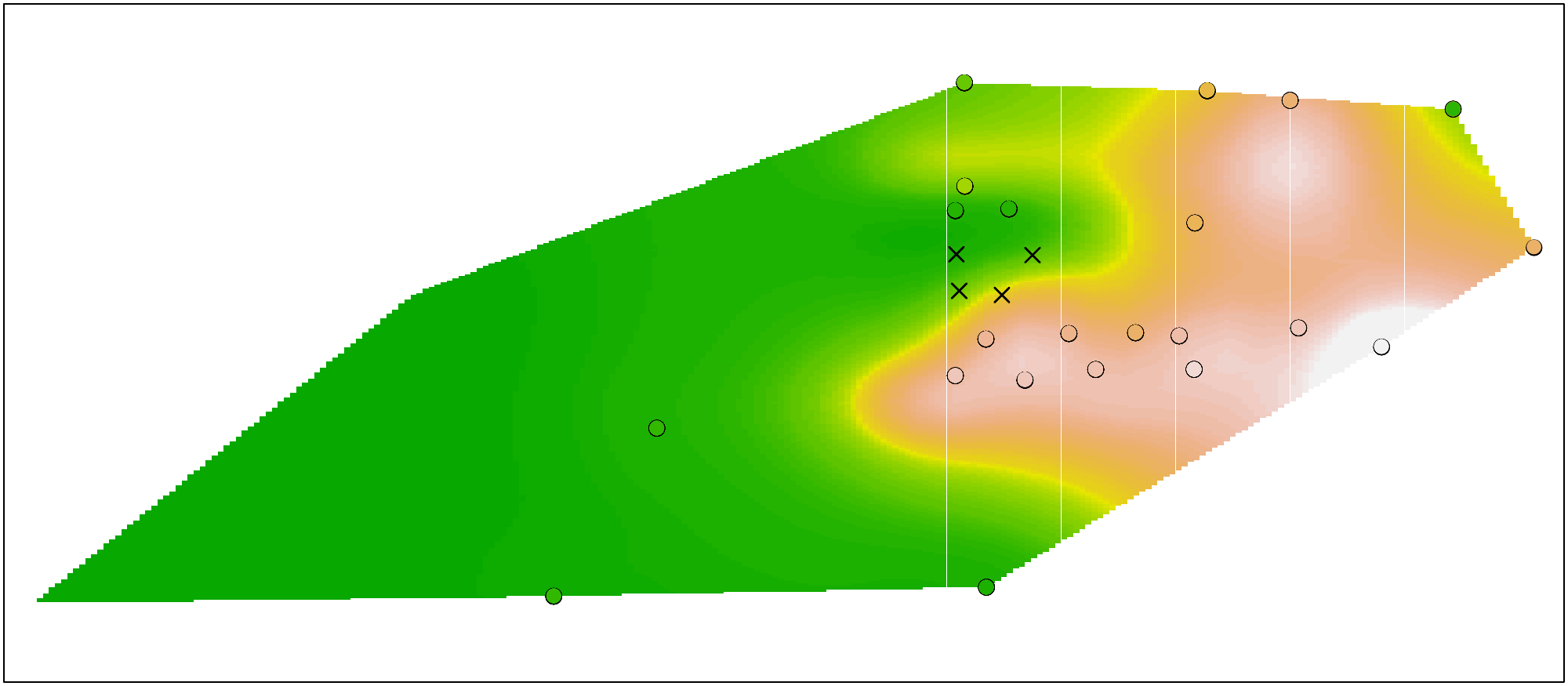}}\\
\subfigure[Well-based cross-validation (all wells)]{\includegraphics[width = 0.45\textwidth]{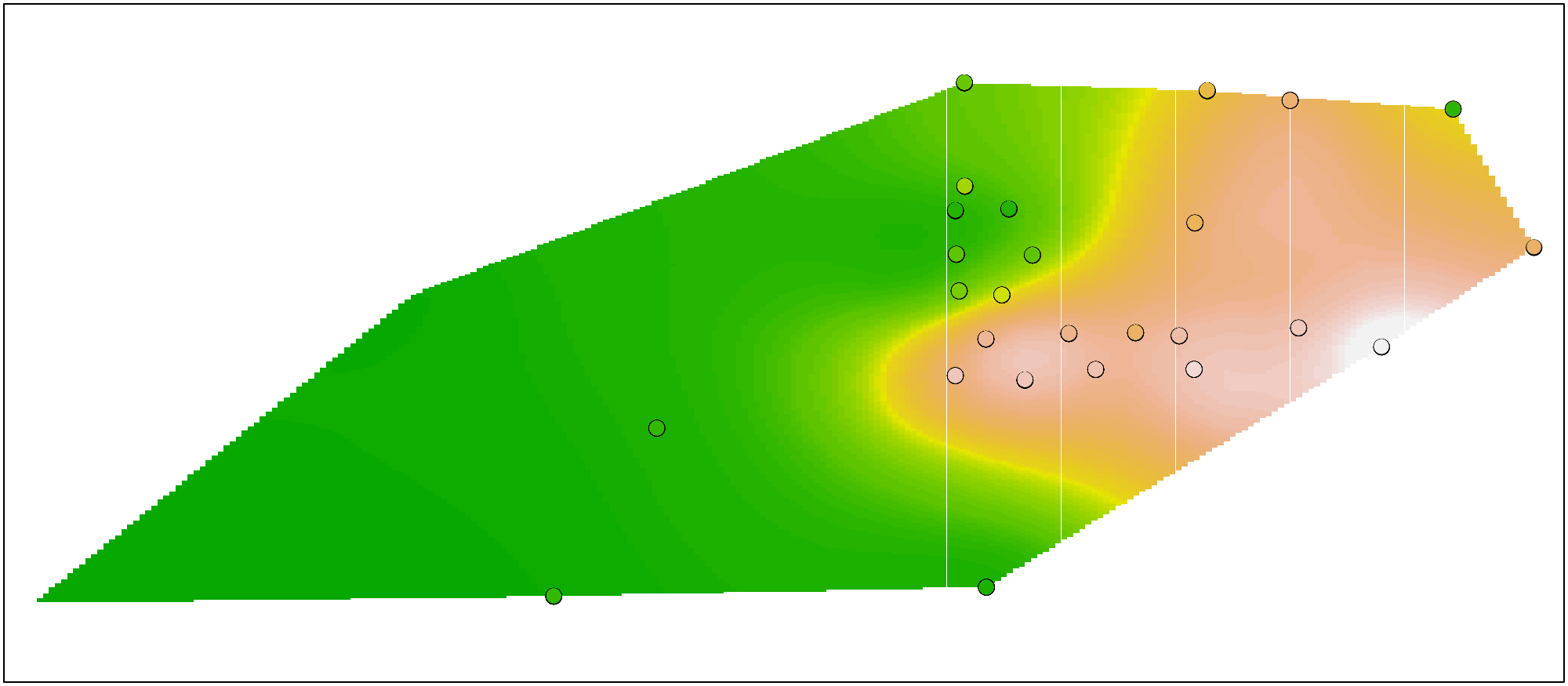}}
\subfigure[Well-based cross-validation (four wells removed)]{\includegraphics[width = 0.45\textwidth]{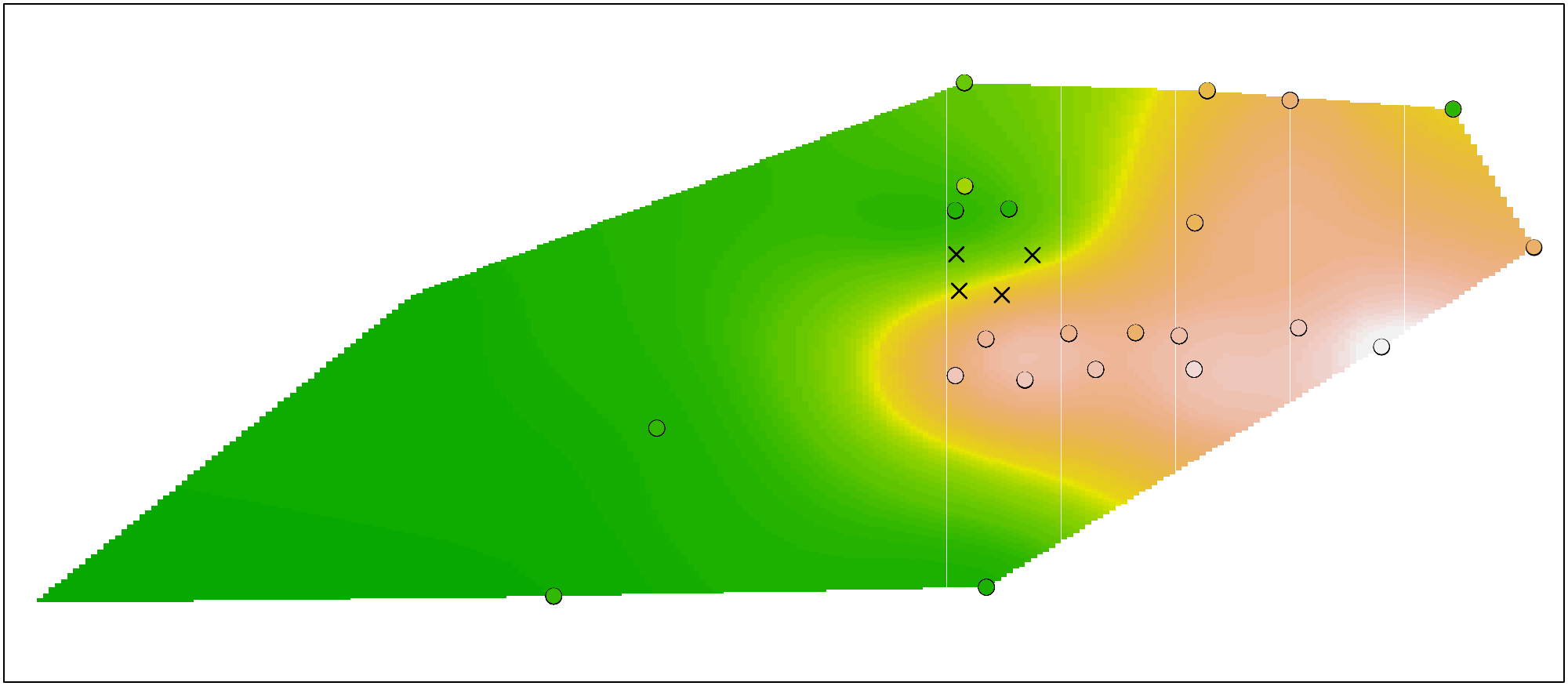}}\\
\subfigure[{\sc bic} (all wells)]{\includegraphics[width = 0.45\textwidth]{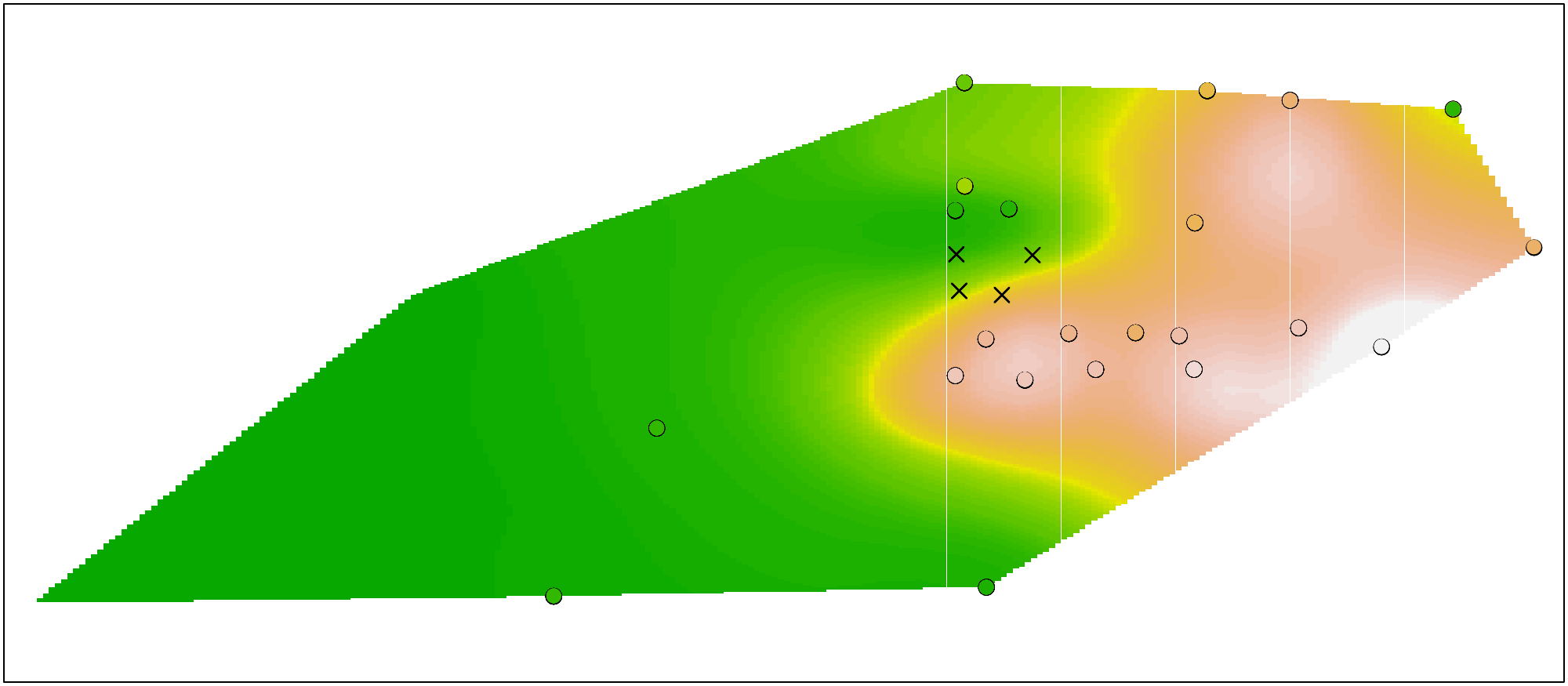}} 
\subfigure[{\sc bic} (four wells removed)]{\includegraphics[width = 0.45\textwidth]{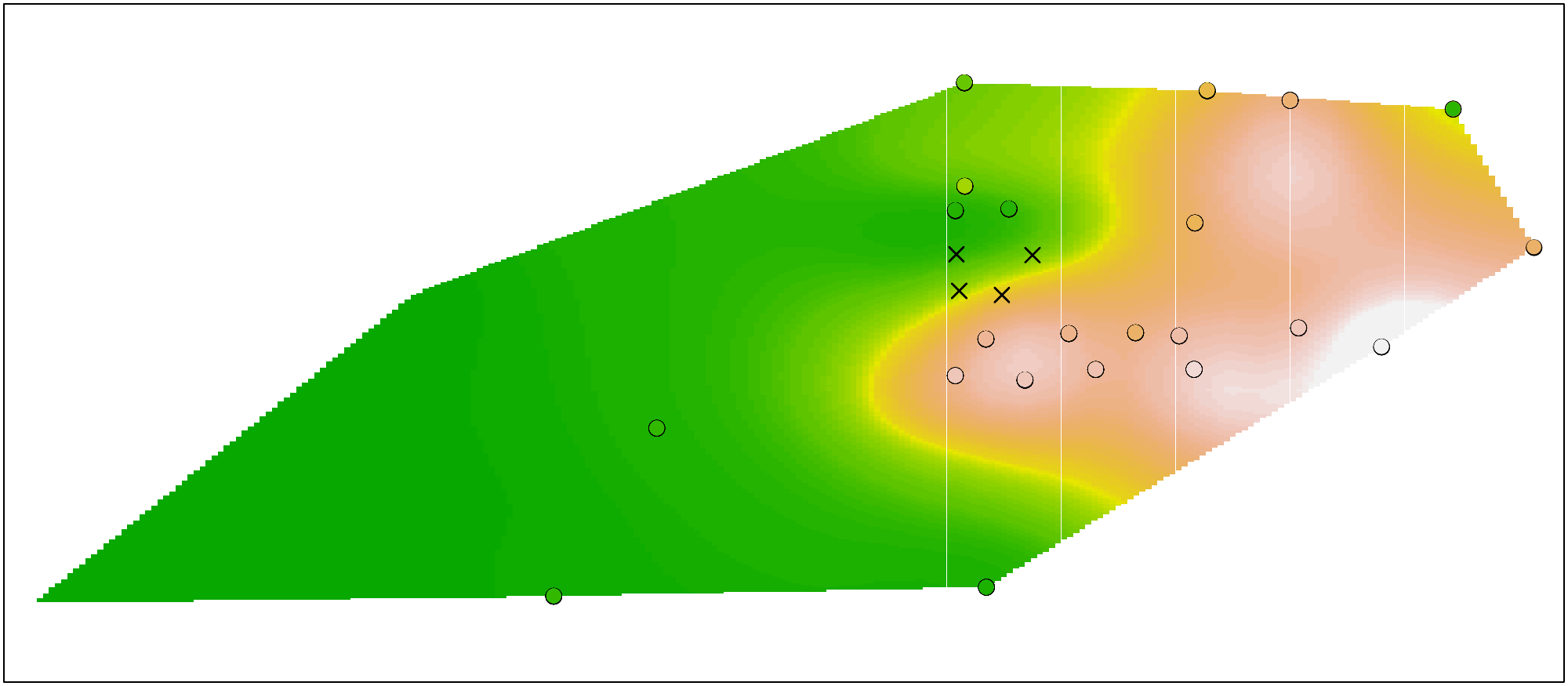}}\\
\subfigure[Bayesian maximum-a-posteriori estimate (all wells)]{\includegraphics[width = 0.45\textwidth]{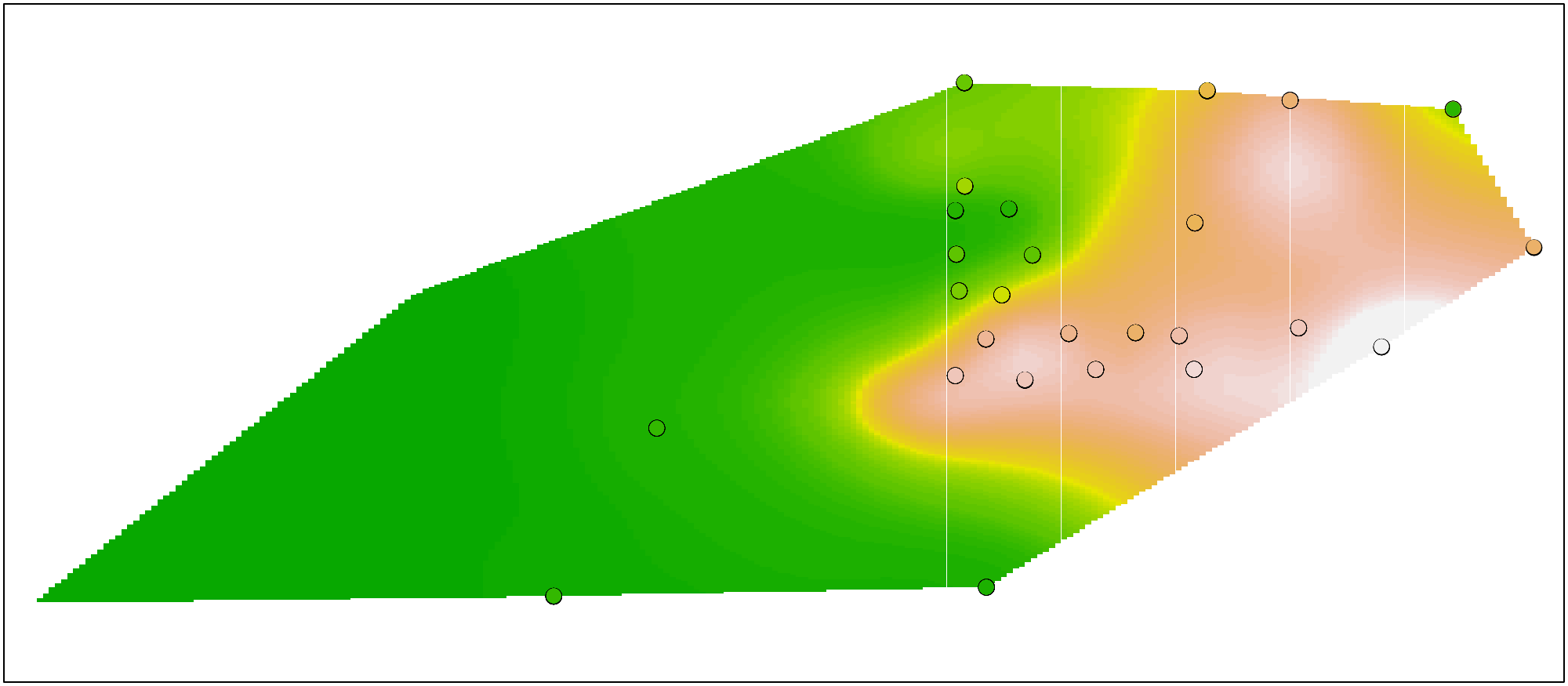}} 
\subfigure[Bayesian maximum-a-posteriori estimate (four wells removed)]{\includegraphics[width = 0.45\textwidth]{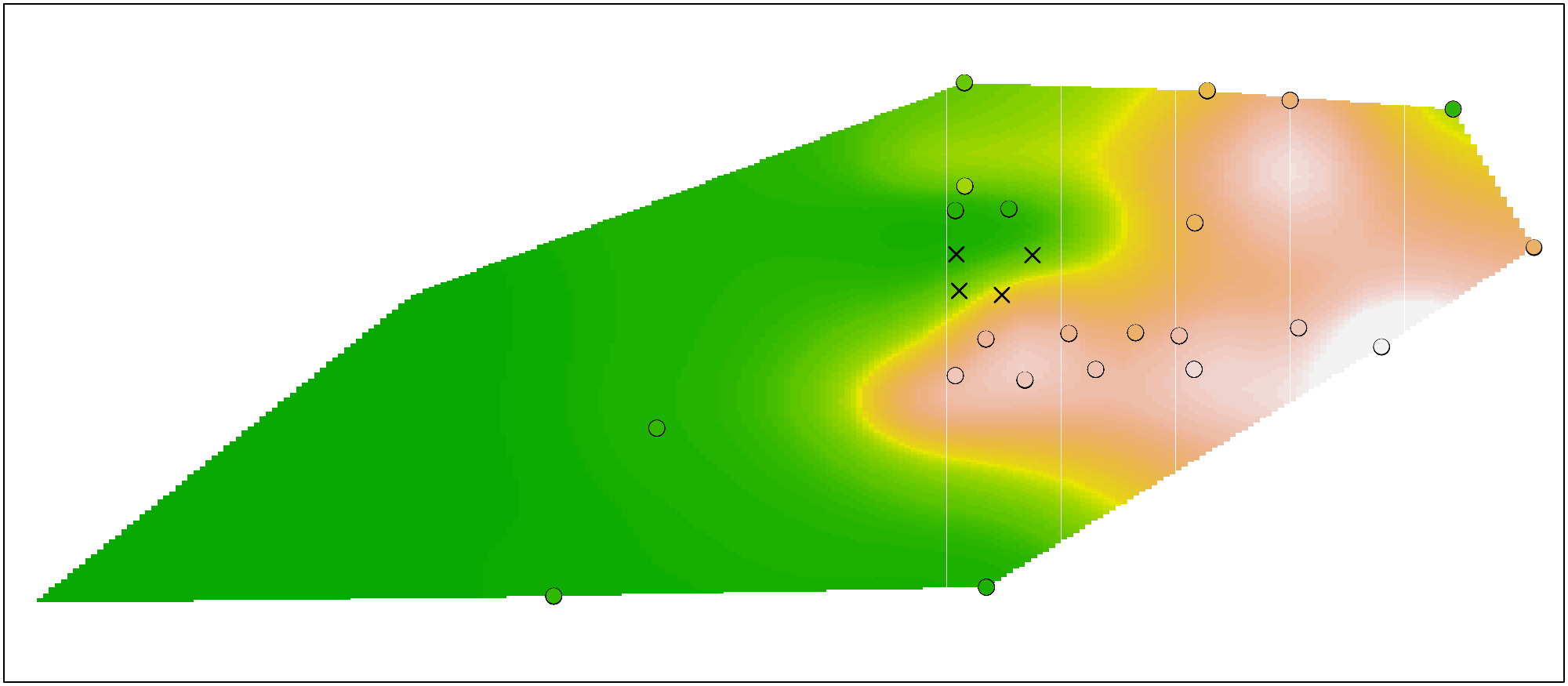}}\\
&  \parbox{0.06\textwidth}{
      \vspace*{-145mm}
      \includegraphics[width = 0.06\textwidth]{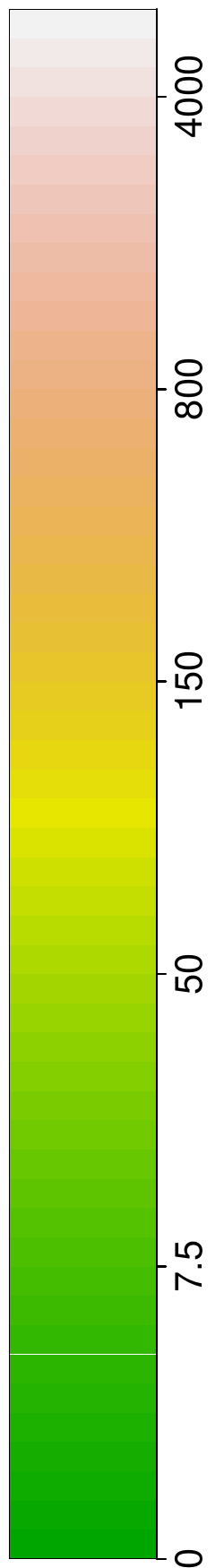}}
\end{tabular}
}
\caption{Predictions of the concentration (in $\mu g/\ell$) for the benzene data on one particular day (see section \ref{subsec:benzene}) using the penalisation parameter chosen by optimizing the different criteria, as well as fully Bayesian model averaging. The left column was obtained by using all wells, the right column was obtained after removing four wells (marked by crosses). Using {\sc gcv} or observation-based cross-validation gives results very similar to {\sc aic}c.}
\label{fig:benzene}
\end{figure}

In the groundwater monitoring setting described in Section~\ref{sec:introduction}, the choice of the degree of smoothing is particularly important, as it needs to be implemented in an unsupervised, automatic setting.  The panels of Figure~\ref{fig:benzene} show the predicted concentrations at a single time snapshot from a dataset of benzene measurements, using a p-spline spatiotemporal model.  The top-left panel shows the effects of using {\sc aic}c in selecting the smoothness of the estimate.  A troubling feature is that there are areas of high predicted values which are not well supported by the observed data (`ballooning'). The top-right panel shows the results when four wells are omitted. Even though the four omitted wells recorded low concentrations, omitting them causes the unsupported peaks of the posterior distribution to disappear, indicating strong sensitivity of the results to particular observations.  A further source of concern is that removing these wells changes the predictions in the immediate vicinity of the wells less than predictions further away.  {\sc gcv} and observation-based cross-validation suffer from the same problem. The results obtained by a Bayesian approach (shown in the bottom two rows) are much less sensitive to the removal of the four wells. The results for cross-validation depend on whether the well structure is used when omitting observations. This will be discussed in more detail in sections \ref{sec:sim} and \ref{sec:groundwater}. The single most important factor which causes this `ballooning' is the design used for collecting the data. However this design is typically imposed by external constraints and thus cannot be changed.

Another key contributing factor to the `ballooning' problems is the choice of the number of basis functions used. The computational advantage of using splines over other methods such a kriging depends strongly on the number of basis functions. In almost all practical applications, this number is chosen by finding a reasonable compromise between run time and memory usage. This is not a problematic issue for one-dimensional or two-dimensional data where a moderate number of basis functions (say around 25) presents little computational challenge. For spatiotemporal data however, using 25 basis functions for each dimension would result in having to perform expensive matrix operations on a $15,625\times15,625$ matrix, which in the context discussed here is not a realistic option.

In the context of spatiotemporal models, a low number of basis functions is usually chosen and this raises the question of whether it is also necessary to use a penalty to induce smoothness, given that the low number of basis functions already offers some protection against over-fitting.  The results discussed in sections \ref{sec:sim} and \ref{sec:groundwater}, however, show that using a penalty is especially important in this case, as it helps to prevent `ballooning'. 

The methods of selecting smoothness and the number of basis functions used are not the only other factors in the appearance of this `ballooning' effect of high and low values.  The use of a second-order smoothness penalty encourages the appearance of linear sections, especially if there is a gap in the data, as these linear sections do not attract any penalty.  Depending on the data around the gap, this can lead to a high peak or deep valley with little support from neighbouring observations.  Using a first-order smoothness penalty typically lessens the problem to some extent, but is, on its own, often not enough to avoid `ballooning' altogether. Choosing a different transformation of the response can, depending on the data and the transform used, also have an effect. 

There is a vast literature on the selection of the smoothing parameter for spline-based models in order to avoid over-fitting or under-fitting. The focus of this work is on the problem of `ballooning', which is a different phenomenon.

\section{Bayesian spatiotemporal smoothing}
\label{sec:bayesian}

This section sets out how a Bayesian framework can be used to select the smoothing parameter. The starting point in the classical Bayesian Linear Model formulation \citep[see e.g. ][]{denison-2002-book} was adapted to the spatiotemporal setting.  If the model determined by a particular value of the penalisation parameter
$\lambda$ is denoted by $M_\lambda$ then, using the model described earlier, the likelihood function is derived from
$$
   \B Y | \B\alpha,\sigma^2,M_{\lambda} \sim 
                            \mathcal{N} \left( \B B \B \alpha, \sigma^2 \B I_n \right)
$$
with $\B{Y} \in  \mathbb{R}^n, \B B \in \mathbb{R}^{n\times m}$ and $\B\alpha \in
\mathbb{R}^m$.
For a fixed value of smoothing parameter, a conjugate prior for the parameters $\B\alpha,\sigma^2$ is the normal-inverse gamma, i.e. 
$$
   \boldsymbol\alpha,\sigma^2|M_{\lambda} \sim 
         \mathcal{NIG} \left(\boldsymbol0, (\lambda\B{D'D})^{-1}, a, b \right) .
$$
In sections \ref{sec:sim} and \ref{sec:groundwater} we use $a = b = 0.0001$ to acknowledge the uncertain prior information on the parameter $\sigma^2$.

The posterior distribution of the penalisation parameter $\lambda$ can be shown to be
$$
   f_{M_\lambda|\B{Y}}\propto \
      \lambda^{\mathrm{rank}(\Db'\Db)/2} \
      \frac{| \B{B}'\B{B} + \lambda\B{D}'\B{D} |^{-1/2}}
             {\bigg\{2b + \B{y'}\Big[\B{I_n}- \B{B}(\B{B}'\B{B} + \lambda\B{D}'\B{D})^{-1}\B{B}' \Big]\B{y}  \bigg\}^{a+ n/2}  } 
                   \  f_{M_\lambda}^{\textrm{prior}} .
$$
This is a special case of model comparison of Bayesian linear models, here using $\lambda$ as the model index.  For a general result see, for example, \citet[equation 2.24, gives the general principle for the comparison of Bayes factors]{denison-2002-book}.  One difficulty in the present context is the degenerate nature of the prior for $\alphab$, expressed in the rank deficiency of the differencing matrix $\Db$.  This can be handled by use of an additional ridge penalty which gives the matrix $\Db$ full rank.  The posterior distribution shown above can be obtained by considering the limit as this ridge penalty goes to $0$. Alternatively, it is possible to decompose the regression coefficients $\alphab$ into two components, one of which has a flat improper prior and the other a proper Gaussian prior.  This approach will be discussed in more detail in the appendix.

The main difference between this approach and the random-effects formulation of \citet{ruppert-2003-book} is that both the regression coefficients and the variance are handled in a Bayesian way, resulting in a fully Bayesian model.

In sections \ref{sec:sim} and \ref{sec:groundwater} a non-informative improper uniform prior is used for $f_{M_\lambda}$.  The value of $\lambda$ which maximises this posterior density, known as the {\sc map} (\textit{maximum a posteriori}) value, is then adopted for penalisation. Additionally, a fully Bayesian approach, numerically integrating out the penalty parameter $\lambda$, is considered.

\section{Computational issues}
\label{sec:linear-algebra}

Computation of the {\sc map} distribution requires the determinant of the posterior covariance matrix of $\alphab$ as well as the posterior residual sum of squares, which in turn requires computation of the penalised least-squares estimator minimising
the objective function (\ref{eqn:objective}). The penalised least-squares estimator and the determinant have to be recomputed for every value of $\lambda$ under consideration.  This requires matrix operations which are $O(p^3)$, where $p$ is the number of regression parameters.  For a spatiotemporal penalised spline model using $p_0$  basis functions in each dimension, we have $p \sim p_0^3$, thus the cost of the matrix operations is $O(p_0^9)$

One can rewrite the problem \citep[see e.g.][]{wood-2000-jrssb,ruppert-2003-book} in such a way that the expensive linear algebra operations can be performed independently of $\lambda$, and only $O(p^2)$ operations have to performed for every value of $\lambda$.  This allows the {\sc map} solution to be computed much more efficiently.

Both the matrix of basis functions $\Bb$ and the differencing matrix $\Db$ are sparse.  Exploiting this sparseness allows further improvement in computational efficiency.  However, for a trivariate p-spline problem with a moderate number of knots, the matrix $\Bb$ is much more dense than the matrix $\Db$.  For example, if 10 basis functions are used for each dimension, roughly 14\% of the entries of $\Bb$ are non-zero, whereas less than 0.3\% of the entries of $\Db$ are non-zero.
Sparsity can therefore be exploited most effectively by initially working only on the matrix $\Db$.  The approaches set out by \citet{wood-2000-jrssb} and \citet{ruppert-2003-book} start by manipulating the matrix $\Bb$ which is less sparse than $\Db$.  The approach set out in detail in the appendix is loosely based on the method described by \citet{Elden-1977}.  The core idea is to exploit the sparseness of the matrices for almost all matrix operations. 
However, each matrix decomposition creates `in-fill' and so the matrices become increasingly dense.  Only the final step, a singular value decomposition, is  computed using dense methods.  This implies that exploiting the sparsity of the design matrix and penalty matrix will not allow the implementation to become much more than twice as fast as the corresponding dense methods.  However, in the context of this work, where the aim is to obtain results within less than a minute, this offers a significant increase in speed.

\section{Simulation study}
\label{sec:sim}

\begin{figure}
\begin{tabular}{ll}
   \subfigure[Groundwater levels and flow used in the simulation]{\includegraphics[width = 0.45\textwidth]{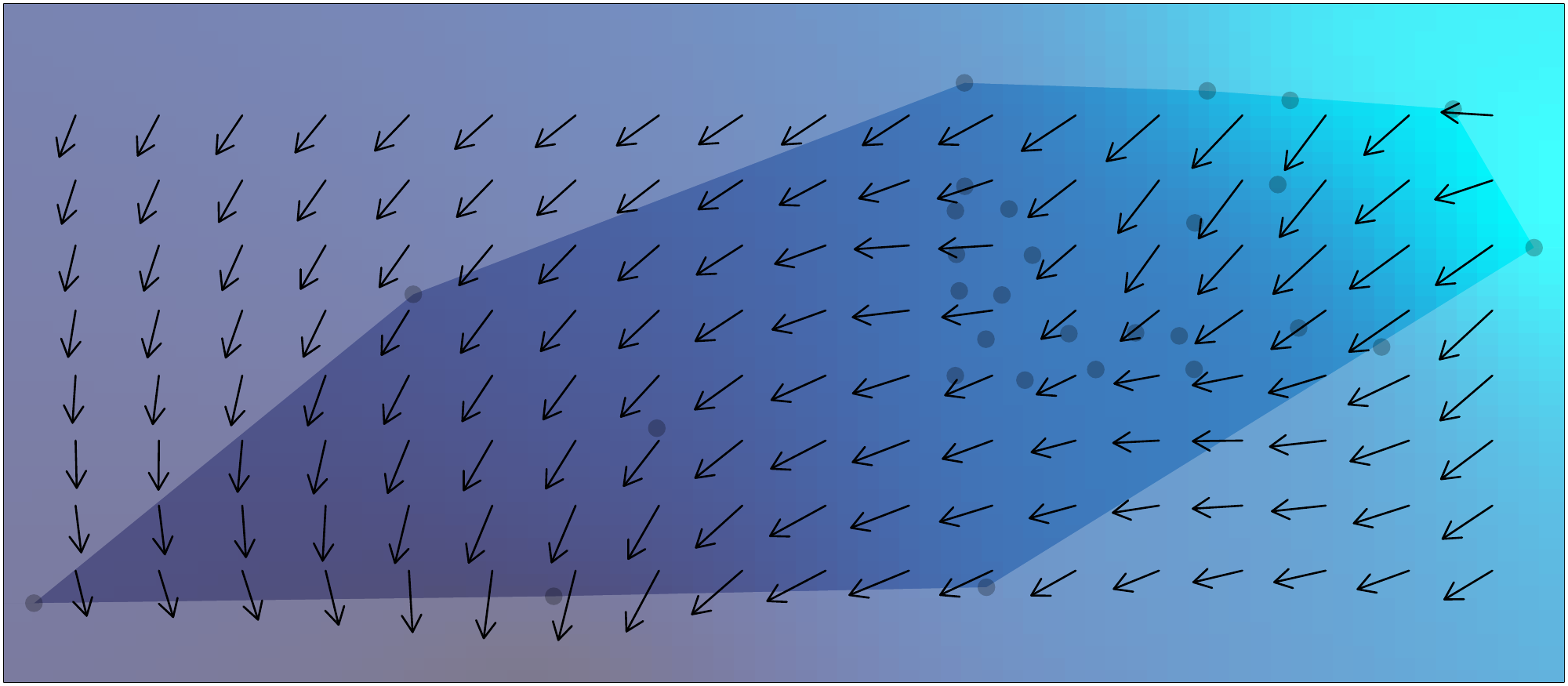}}
   \subfigure[Initial condition of the concentrations at time $t=0$]{\includegraphics[width = 0.45\textwidth]{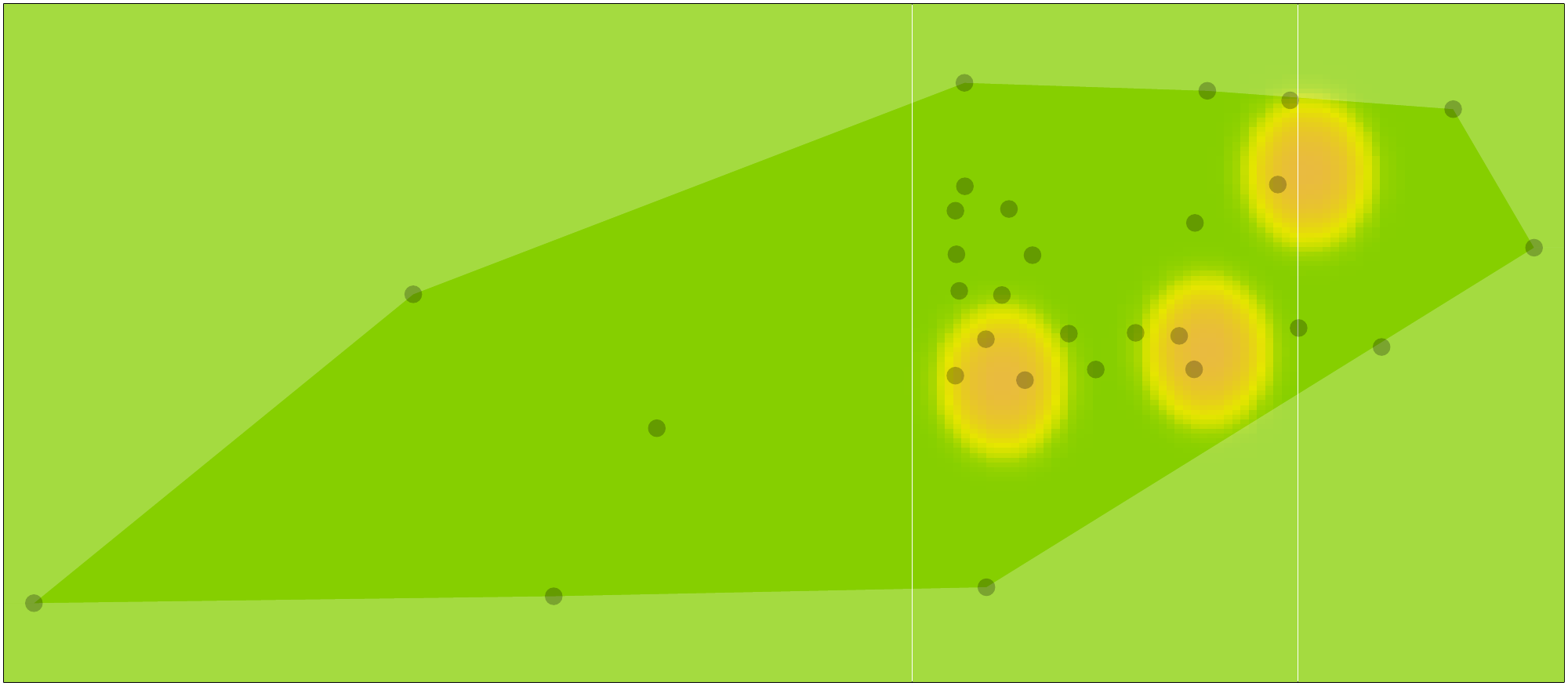}}
&
   \includegraphics[width = 0.028\textwidth]{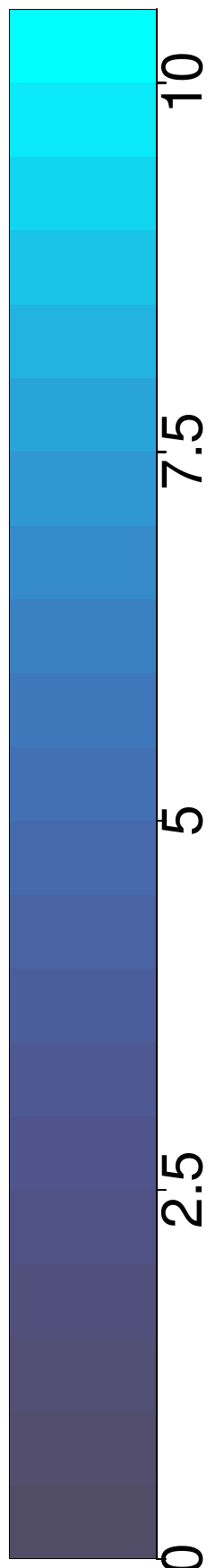}
\\
  \subfigure[Simulated concentrations at time $t=0.25$]{\includegraphics[width = 0.45\textwidth]{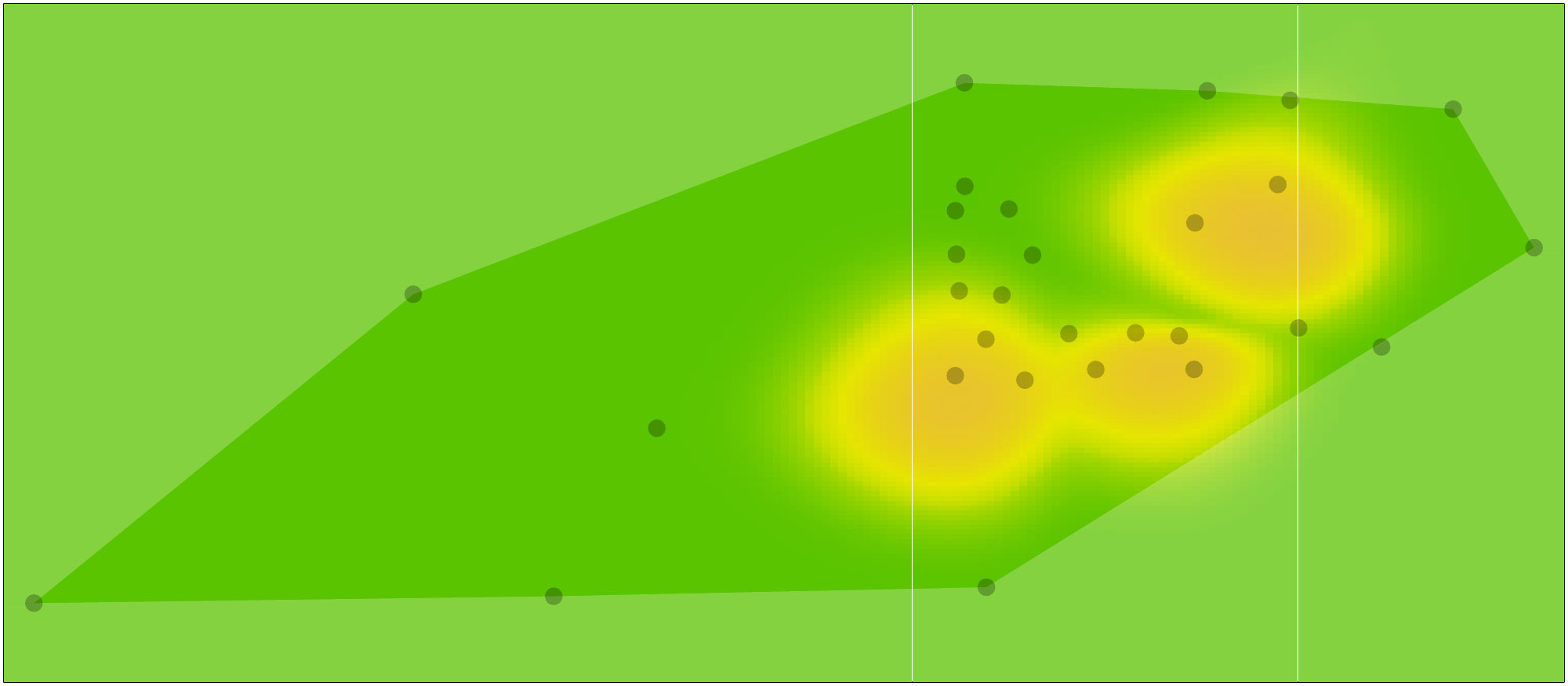}}
  \subfigure[Simulated concentrations at time $t=0.5$]{\includegraphics[width = 0.45\textwidth]{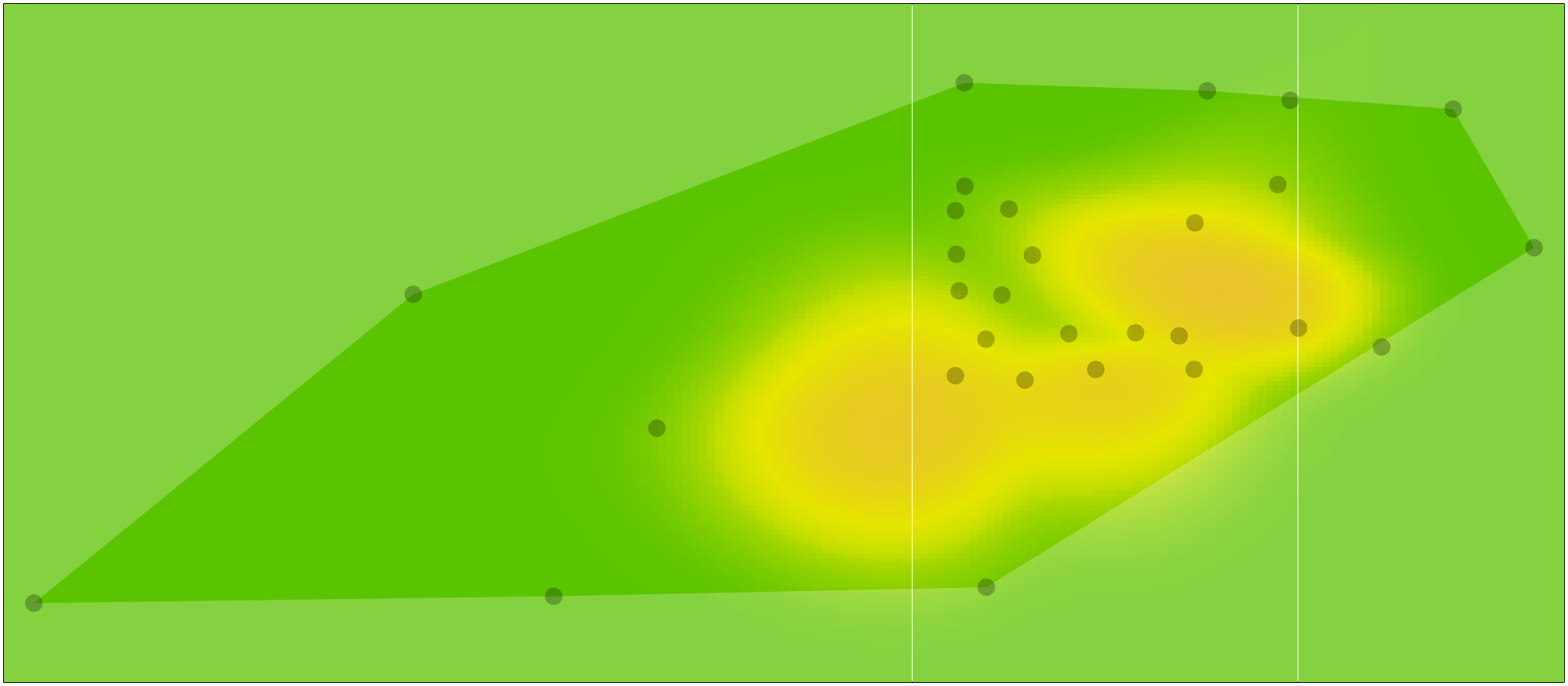}}
&
   \includegraphics[width = 0.028\textwidth]{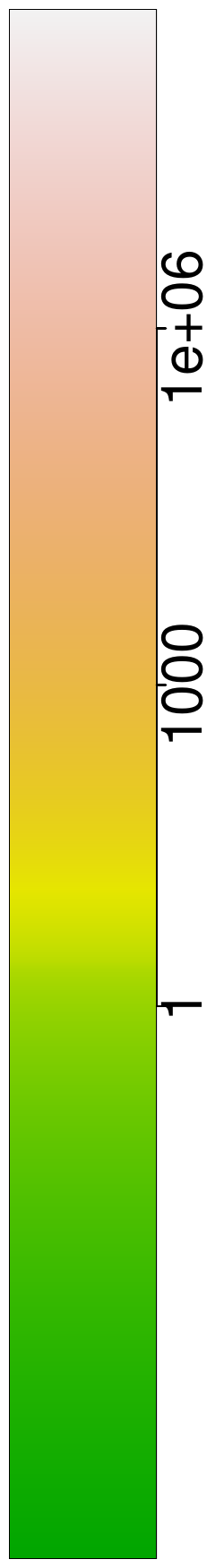}
\end{tabular}
\caption{Flow model, initial concentrations and simulated concentrations for $t\in\{0.25,0.5\}$ used in the simulation study. The simulated concentrations for $t=0.7$ are shown in figure \ref{fig:sim}(a).}
\label{fig:sim0}
\end{figure}

In this section, a simple simulation study is used to compare the different methods of selecting the smoothing parameter in a systematic way. The data are simulated from a highly idealised model for the spread of a solute in water. This is based on the partial differential equation 
$$
\frac{\partial y}{\partial t}=
D\cdot \left (\frac{\partial^2 y}{\partial s_1^2}+\frac{\partial^2 y}{\partial s_2^2}\right) + \psi_1(s_1, s_2) \frac{\partial y}{\partial s_1} + \psi_2(s_1, s_2) \frac{\partial y}{\partial s_2} .
$$
Here $y$ denotes the concentration of the solute, $s_1$ and $s_2$ denote the spatial coordinates and $t\in [0,1]$ denotes time. The first term describes the spread of the solute in the groundwater by diffusion, with the constant $D$ controlling how fast the solute spreads. The two further advection terms describe how the solute is affected by groundwater flow, whose direction and velocity is represented by the functions $\psi_1$ and $\psi_2$.  These functions were chosen to correspond to the observed groundwater levels in the benzene example discussed in section \ref{subsec:benzene}. Figure \ref{fig:sim0}(a) shows the assumed groundwater levels and flow which, in the simulations, are assumed for simplicity to be constant over time. 

\begin{figure}
   \subfigure[Simulated true concentrations (base model at $t=0.7$)]{\includegraphics[width = 0.45\textwidth]{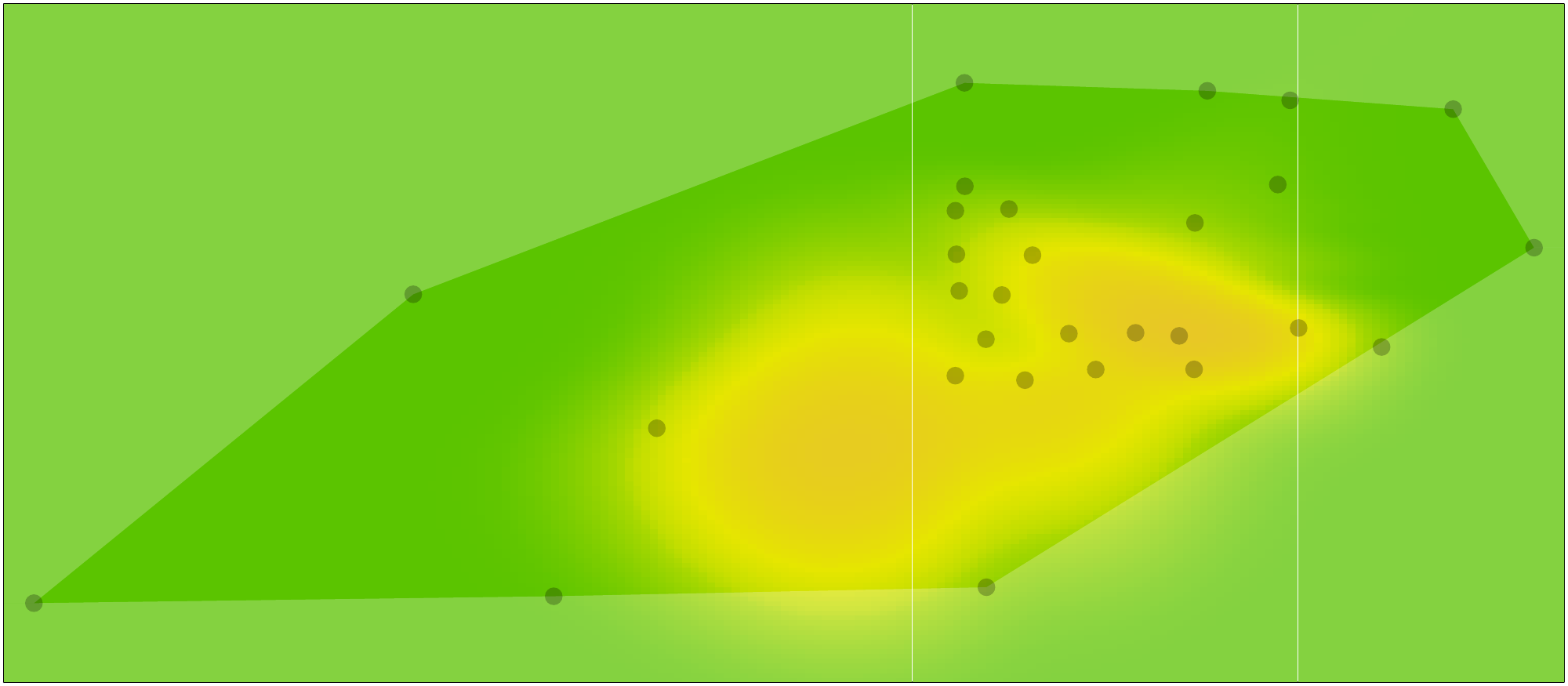}}
   \subfigure[{\sc aic}c]{\includegraphics[width = 0.45\textwidth]{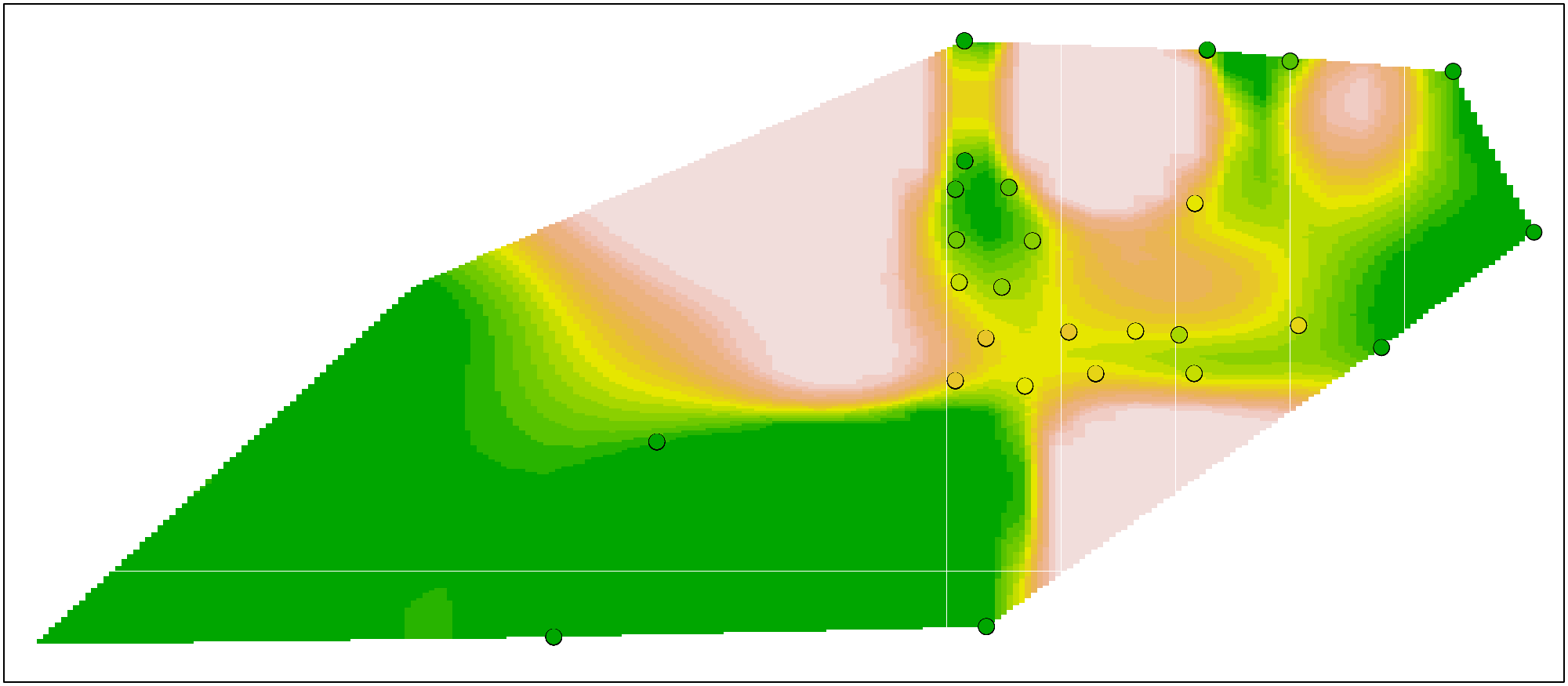}}

   \subfigure[{\sc bic}]{\includegraphics[width = 0.45\textwidth]{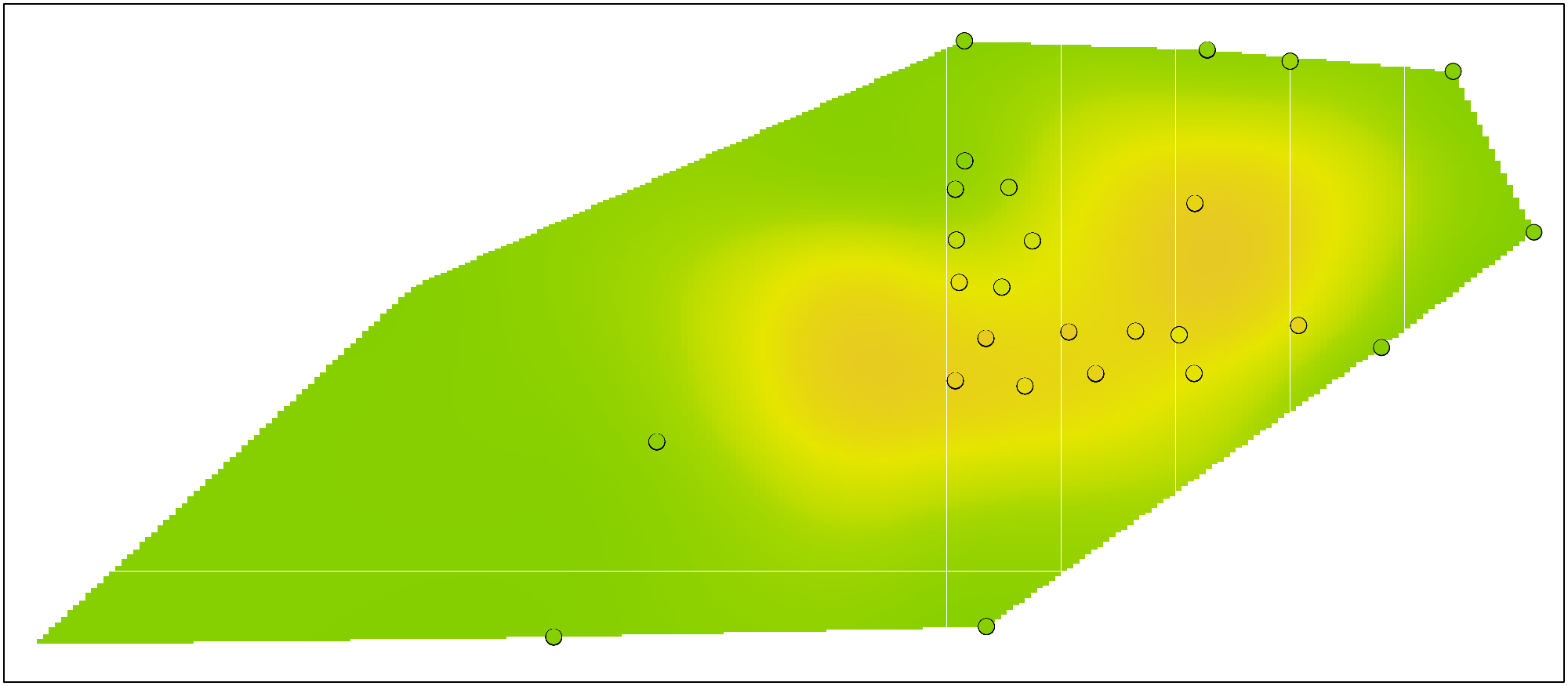}}
   \subfigure[{\sc gcv}]{\includegraphics[width = 0.45\textwidth]{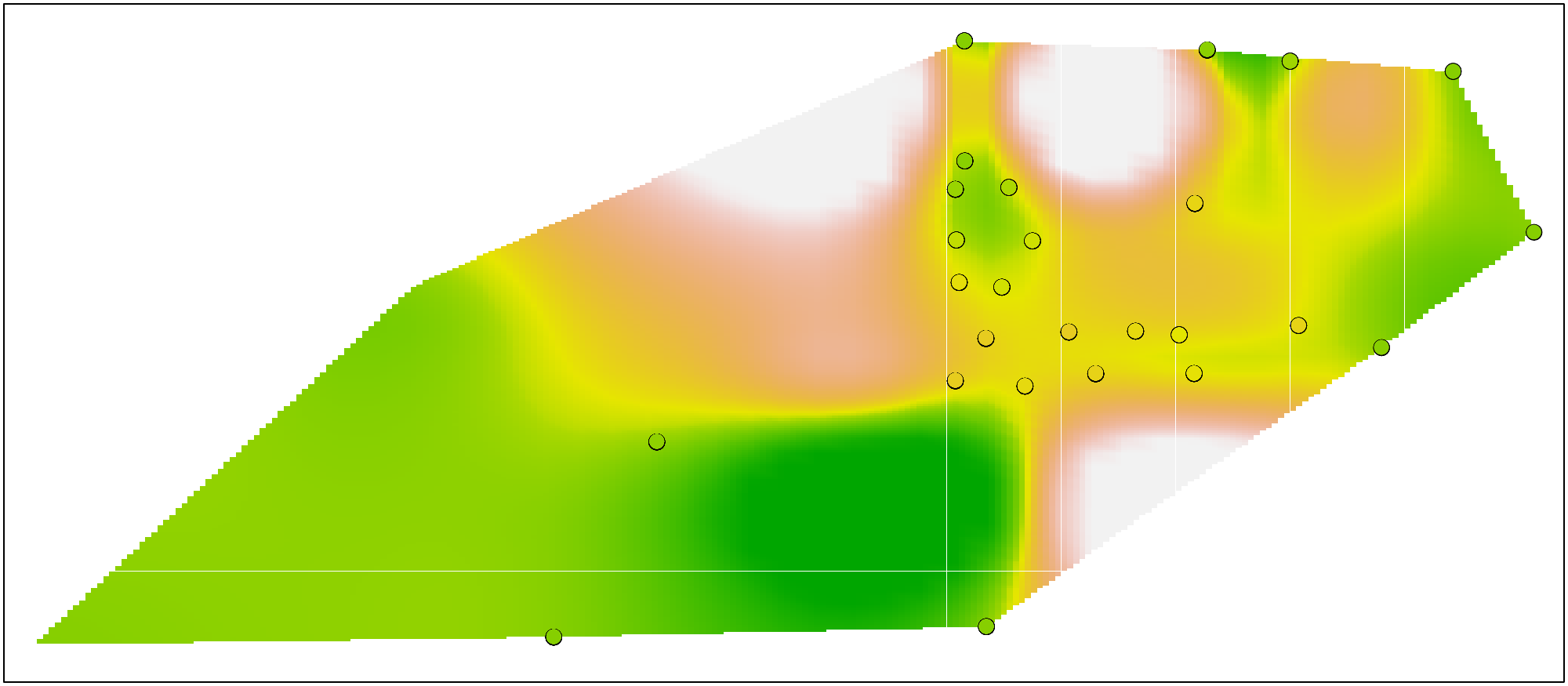}}

   \subfigure[Well-based cross-validation]{\includegraphics[width = 0.45\textwidth]{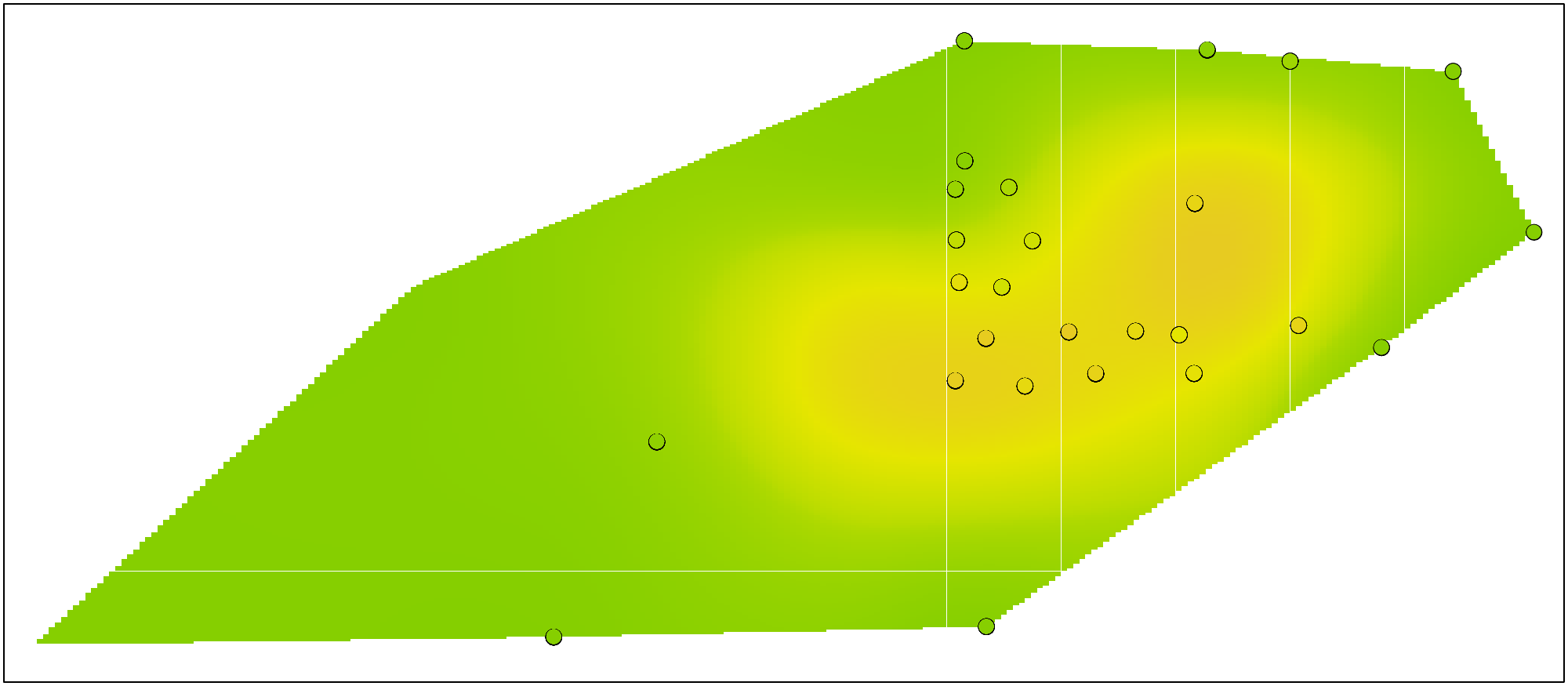}}
   \subfigure[Observation-based cross-validation]{\includegraphics[width = 0.45\textwidth]{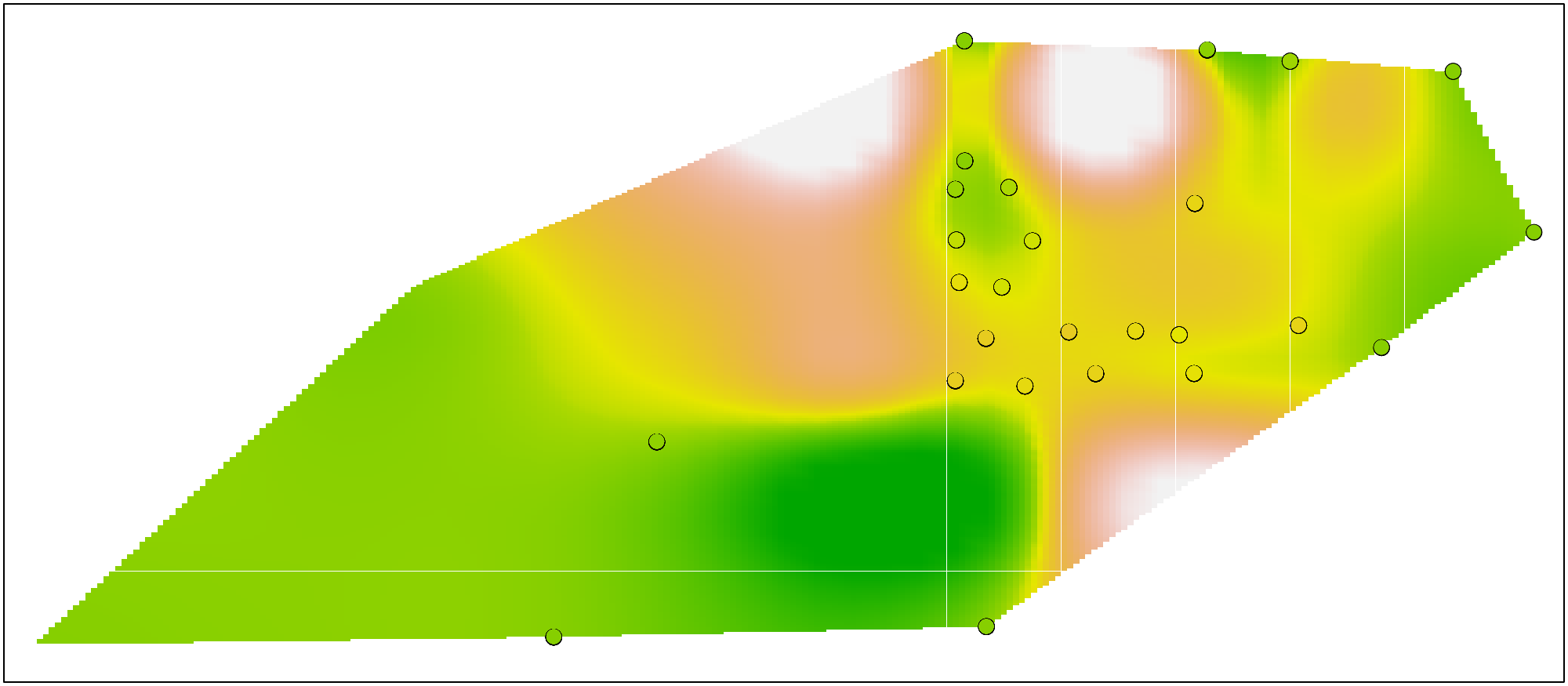}}

   \subfigure[Bayesian maximum-a-posteriori estimate]{\includegraphics[width = 0.45\textwidth]{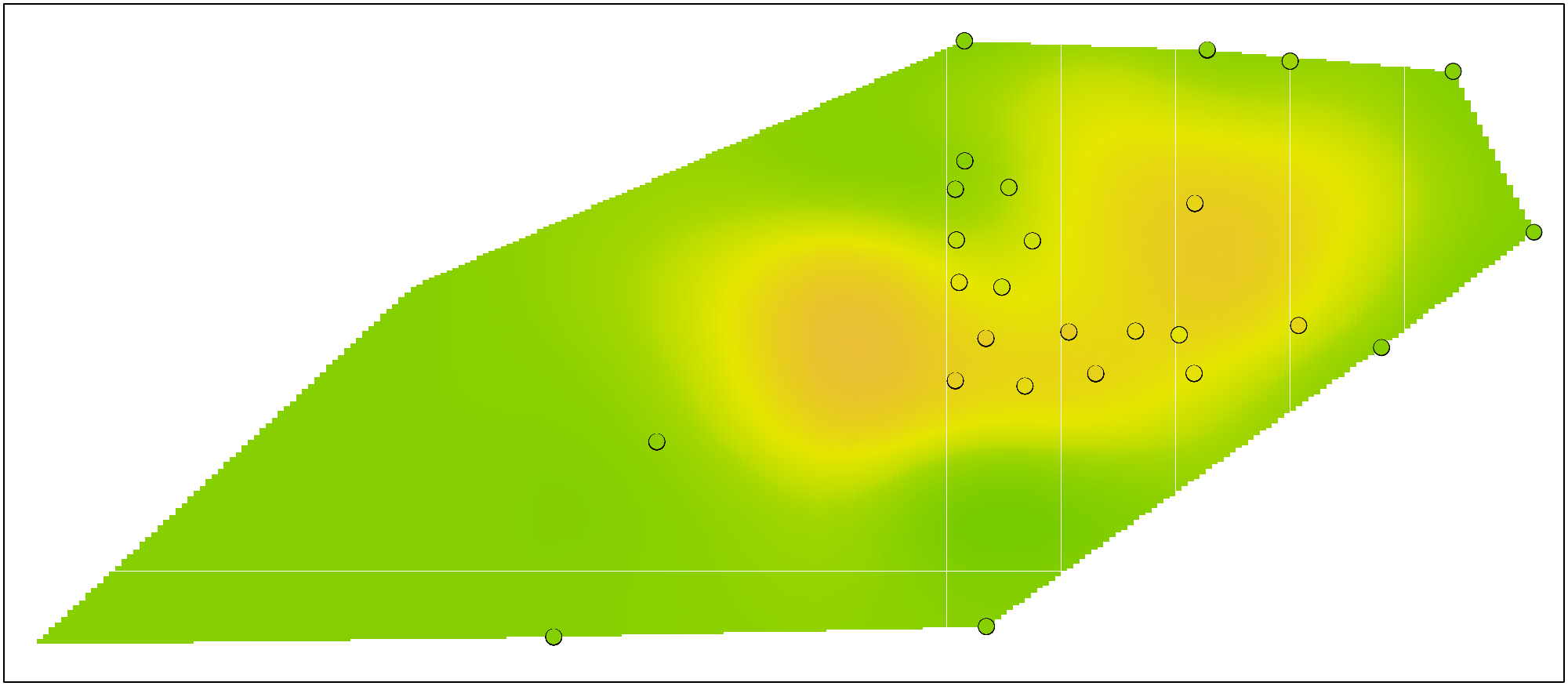}}
   \subfigure[Bayesian model averaging]{\includegraphics[width = 0.45\textwidth]{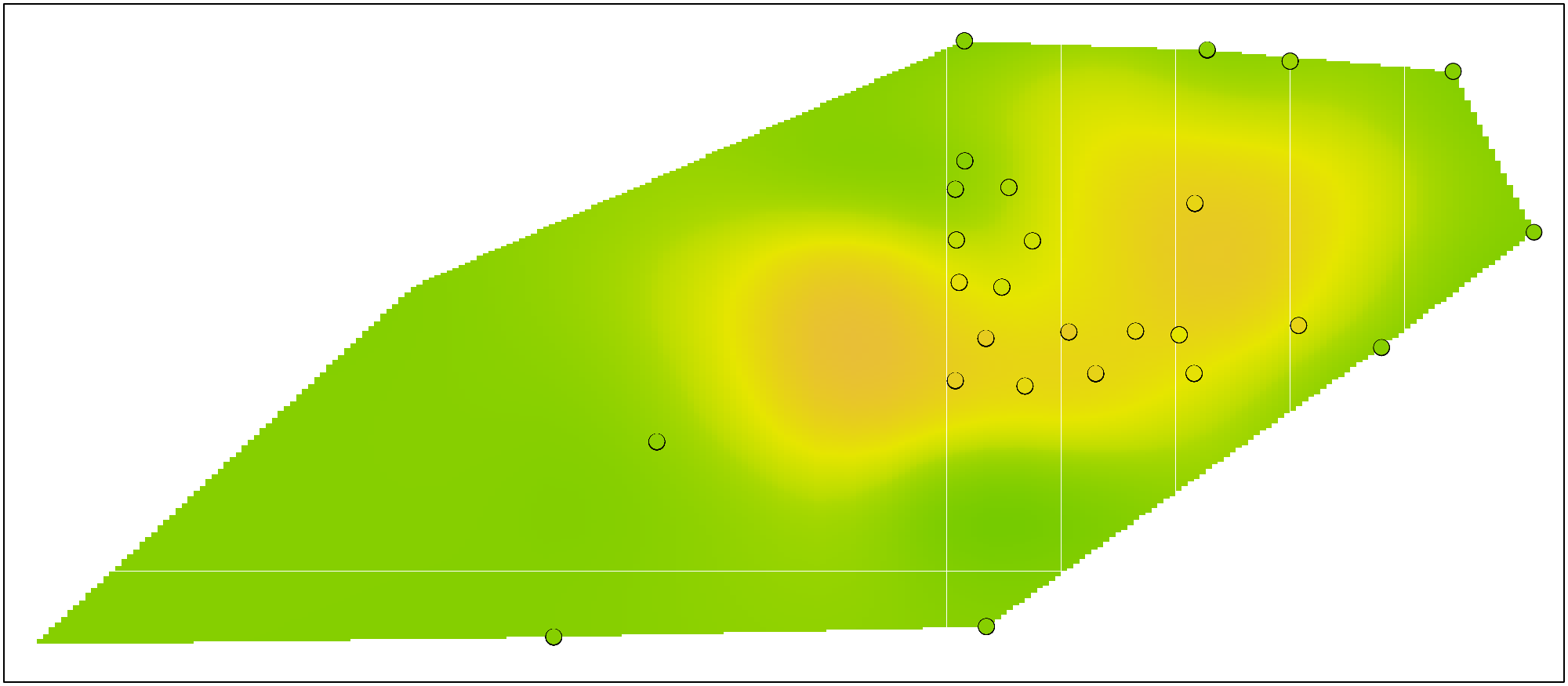}}

\caption{Simulated true model (top left) as well as predictions obtained in one iteration of the simulation at time $t=0.7$ using the wells from scenario 1. Each panel corresponds to the use of a different criterion for selecting the smoothing parameter.}
\label{fig:sim}
\end{figure}

The assumed initial spread of the solute is given in Figure \ref{fig:sim0}(b). Figures \ref{fig:sim0}(c), \ref{fig:sim0}(d) and \ref{fig:sim}(a) show the spread at time $t\in\{0.25,0.5,0.7\}$.  The ``true'' concentrations were obtained by interpolating the numerical solution to the differential equation, computed over a $100\times 100\times 100$ regular grid.  Observed measurement data were generated by multiplicative Gaussian error terms, with standard deviation chosen to give a signal-to-noise ratio on the log-scale of $10:1$. This reflects the fact that measurements of the solutes are usually quite accurate.  A very small value of $0.05$ was used for within-well correlation of the data, while the between-well correlation was assumed to be $0$.  Before the data were analysed they were transformed using the function $\log(y+1)$. The additive term was introduced because the simulations can produce concentrations of exactly $0$.  All model fitting and evaluation was performed on the transformed scale.

A p-spline model using second order basis functions and a first order penalty was used with $14$ basis functions for easting, $8$ for northing and $5$ for time.  The different number of basis functions for space match the different extents of the monitored region in easting and northing in the guiding example, while the reduced number of basis functions for time was chosen to reflect the fact that concentrations vary more quickly in space than in time.  Addressing these issues through the basis functions allows a single smoothing parameter to be used in the model.  Where little \textit{a priori} information on solute behaviour is available, a natural default would be to choose a common number of basis function in each dimension.  The overall number of basis functions is deliberately chosen to be rather low to allow fast computations.  Experimentation has shown these numbers of basis functions to be effective from this perspective, in addition to preventing `ballooning' or over-fitting and producing good estimates of the underlying solute patterns.

Three different designs were used. The first scenario uses exactly the same well coordinates and sample dates as the benzene example discussed in section \ref{subsec:benzene}. It consists of $1402$ observations sampled at $29$ well locations. The second scenario uses a much larger number of $280$ randomly placed wells which are sampled much less frequently, resulting in the same number of observations.  The second scenario is a much better design from a statistical point of view but is, of course, much more expensive, as establishing a new well is considerably more costly than collecting a sample from an existing one.  The third scenario uses the same wells as the first scenario, but only has $100$ observations in total, with each well sampled only about four times on average. 

For all three scenarios, the methods were compared using the integrated squared error
$$
\underset{\mathcal{S}}{\int\int\int} (\hat m(s_1, s_2,t) - m(s_1, s_2, t))^2 ds_1 ds_2 dt
$$
The integral was computed numerically over the inside of the convex hull of the observed wells and sampling dates. 

Table \ref{tab:sim} shows the results obtained from 500 replications for all three scenarios. From the table it is immediately clear that no one method outperforms all other methods for all three scenarios.

\begin{figure}
\centerline{\begin{tabular}{ll}
\subfigure[Well scenario 1]{\includegraphics[width=.7\textwidth]{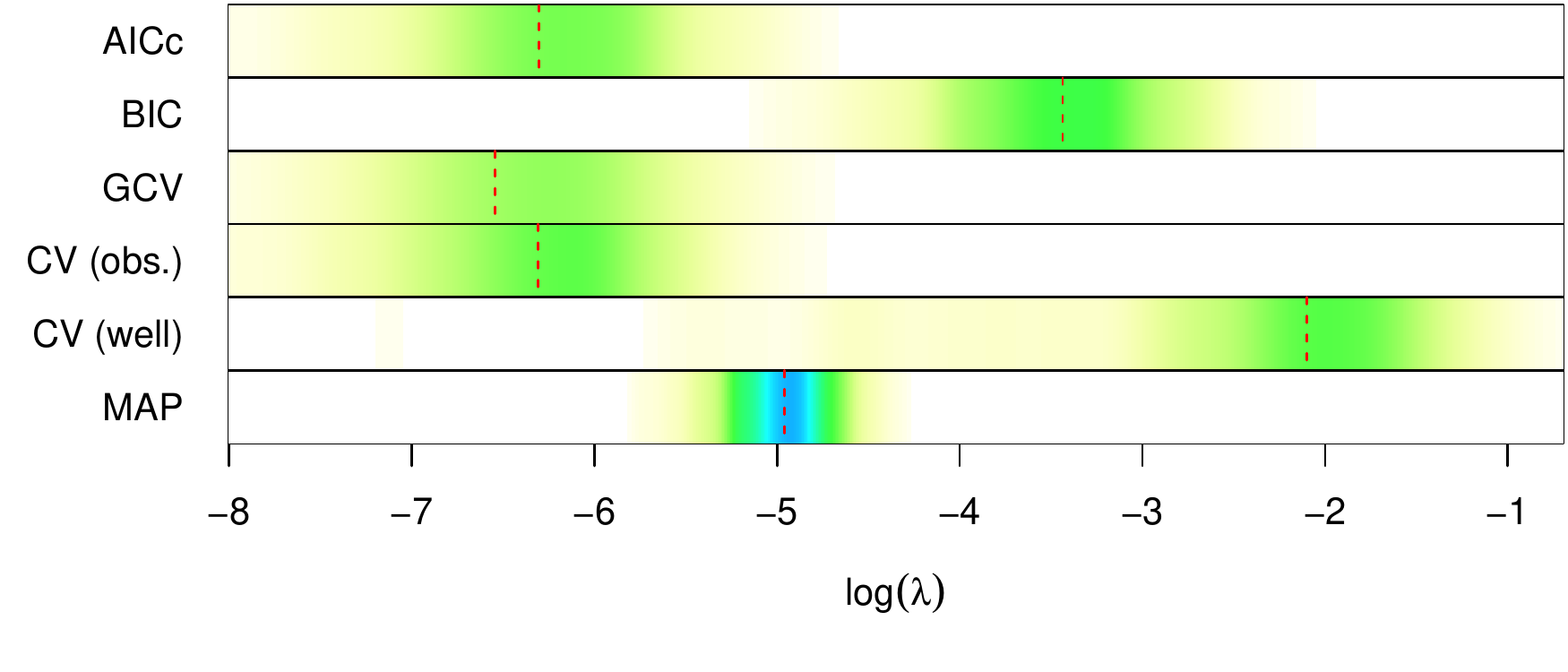}}\\
\subfigure[Well scenario 2]{\includegraphics[width=.7\textwidth]{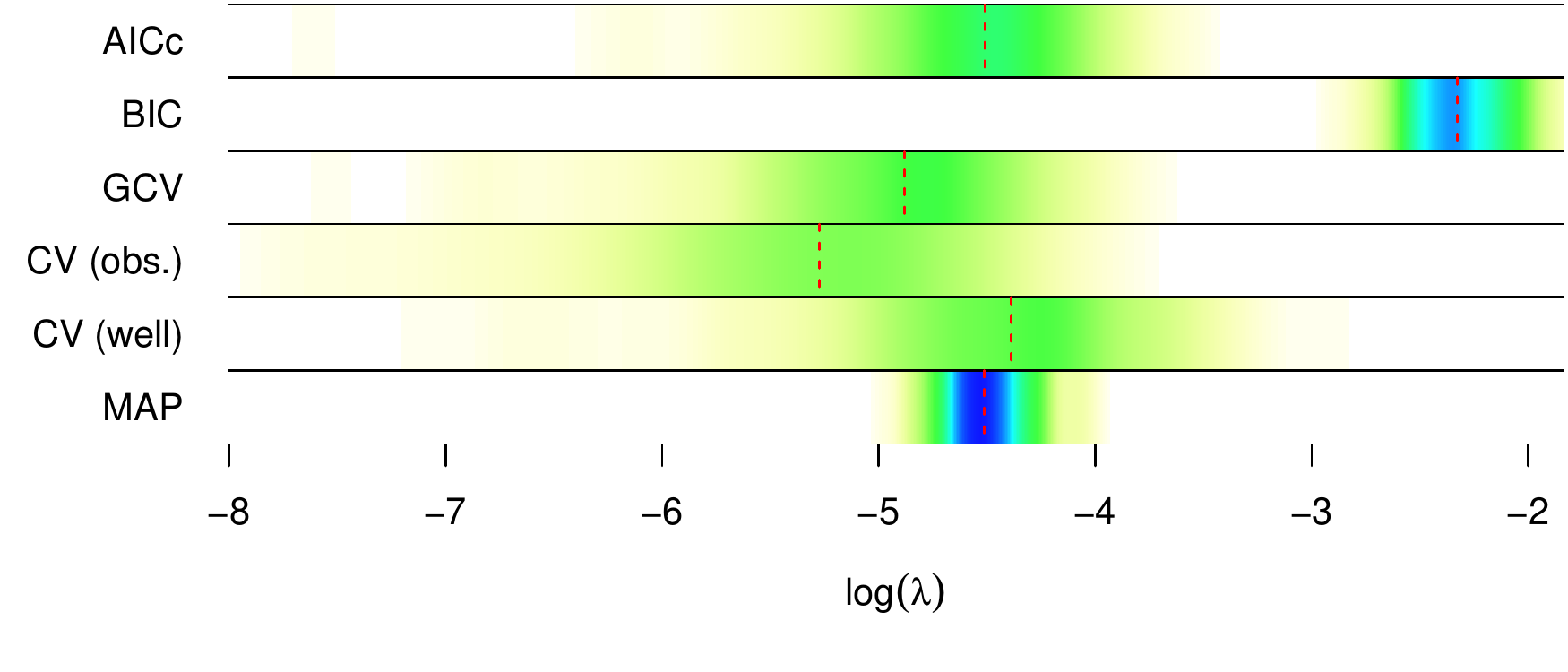}}\\
\subfigure[Well scenario 3]{\includegraphics[width=.7\textwidth]{figures/densitystrip2.pdf}}&
\parbox{0.06\textwidth}{
      \vspace*{-165mm}
      \includegraphics[width = 0.06\textwidth]{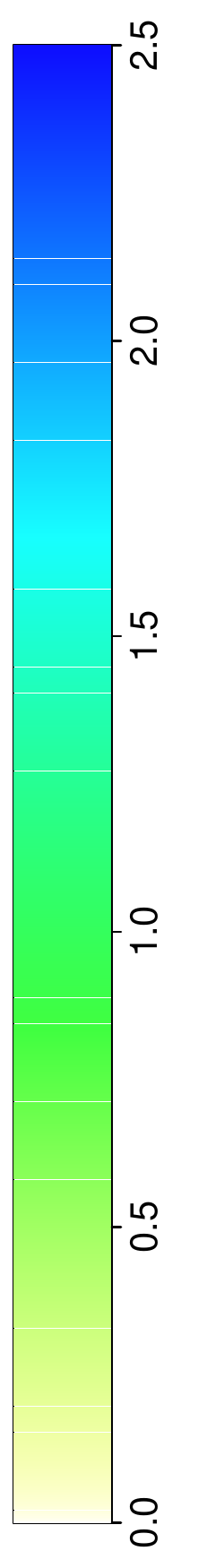}}
\end{tabular}}
\caption{Density strip plots of the smoothing parameters chosen by the different methods for both scenarios. The dashed red line indicates the median.}
\label{fig:densstrip}
\end{figure}

Out of the three scenarios presented, only scenario one is prone to `ballooning'. In this situation {\sc aic}c and {\sc gcv} show poor performance.  Figure \ref{fig:sim}(b)--(h) shows the reason for the poor performance of these, and observation-based cross-validation, as all three lead to severe `ballooning'. The Bayesian approaches ({\sc bic}, {\sc map} and model averaging) give much better performance and Figure \ref{fig:sim} shows no evidence of `ballooning'.  Figure \ref{fig:densstrip} (a) shows density strip plots \citep{densstrip} of the distribution of the smoothing parameter $\lambda$ for each method.  This shows that the Bayesian approaches and well-based cross-validation select values of the smoothing parameter $\lambda$ which are large enough to prevent `ballooning'.  The problems with other methods are caused by values of $\lambda$ which are too low.

Though {\sc bic} performs very well if the focus is on preventing ballooning, it is prone to under-fitting. In the second scenario, which provides the `best' data for estimating the concentrations, {\sc bic} performs significantly worse than the other methods. As Figure \ref{fig:densstrip} (b), shows this is due to selecting a value for $\lambda$ which is too large.

In all three scenarios, the {\sc map} and the fully Bayesian approach give good results, being the best method in the second and the third scenarios.

Cross-validation is, by far, the most computationally demanding method.  In the simulations, and in section~\ref{sec:groundwater}, $10$-fold cross-validation was used.  The results depend on how the cross-validation is carried out.  One option is to remove entire wells rather than single observations and in this case cross-validation favours very large values of the penalty parameter.  In contrast, if the well structure is ignored and individual observations are removed, cross-validation favours very small values of the penalty parameter.  The reason for the difference is that, in this dataset, `ballooning' occurs only in space and not in time.  There is a relatively small number of wells and these are sampled very frequently in time. Omitting observations individually typically does not create gaps in time which are large enough to allow `ballooning' at individual wells.  Cross-validation can therefore address `ballooning' only if a well is omitted entirely. The difference between the two variants is much less pronounced in the second and third scenario. 

\begin{table}
\begin{center}
\begin{tabular}{lrrp{0.1cm}rrp{0.1cm}rr}
\hline Criterion used to&\multicolumn{2}{c}{Scenario 1}&&\multicolumn{2}{c}{Scenario 2}&&\multicolumn{2}{c}{Scenario 3}\\
\cline{2-3}\cline{5-6}\cline{8-9}select smoothness&mean& (std. err.)&&mean&(std. err)&&mean&(std. err)\\\hline
{\sc aic}c&313.326 & (258.319)&& 0.137 & (0.003)&& 0.870 & (0.009)  \\
{\sc gcv}&348.947 & (276.003)&&0.139 & (0.001) && 3.780 & (2.282) \\
obs.-based {\sc cv}&3.552& (0.216)&&0.141 & (0.001)&& 1.023 & (0.016)  \\
well-based {\sc cv}&0.809 & (0.012)&&0.138 & (0.001)&& 0.902 & (0.012)  \\
{\sc bic}&1.028 & (0.013)&&0.179 & (0.001)&& 0.875 & (0.003)  \\
Bayesian {\sc map}&1.662& (0.020)&&0.136 & (0.001)&& 0.863 & (0.005)  \\
Bayesian model avg.&1.639 & (0.019)&&0.136 & (0.001)&& 0.858 & (0.005)  
\\\hline\end{tabular}
\end{center}
\caption{Integrated squared errors of the predictions averaged over the convex hull of the data as well we sum of squared errors at the wells obtained from 500 replications for the three well scenarios.}
\label{tab:sim}
\end{table}

\section{Application to groundwater monitoring}
\label{sec:groundwater}
\subsection{Monitoring of benzene in groundwater}
\label{subsec:benzene}
Benzene (C${}_6$H${}_6$) is a constituent of crude oil and refined petrol, which can have serious adverse health (and ecological) effects if released into the environment.  A release from an underground storage tank system can result in benzene contaminating the groundwater below the storage tank system.  After such releases, networks of wells are set up to monitor possible groundwater contamination.  The contaminant of interest is the concentration of benzene in $\mu g/\ell$ (modelled on a log-scale). The data consists of $1402$ observations, which were obtained from a network of $29$ wells. A $p$-spline model using second-order basis functions ($18$ for easting, $11$ for northing and $7$ for time) and first-order difference penalties was fitted, with the smoothing parameter determined using different criteria. 

\begin{figure}
\centerline{\includegraphics[width = \textwidth]{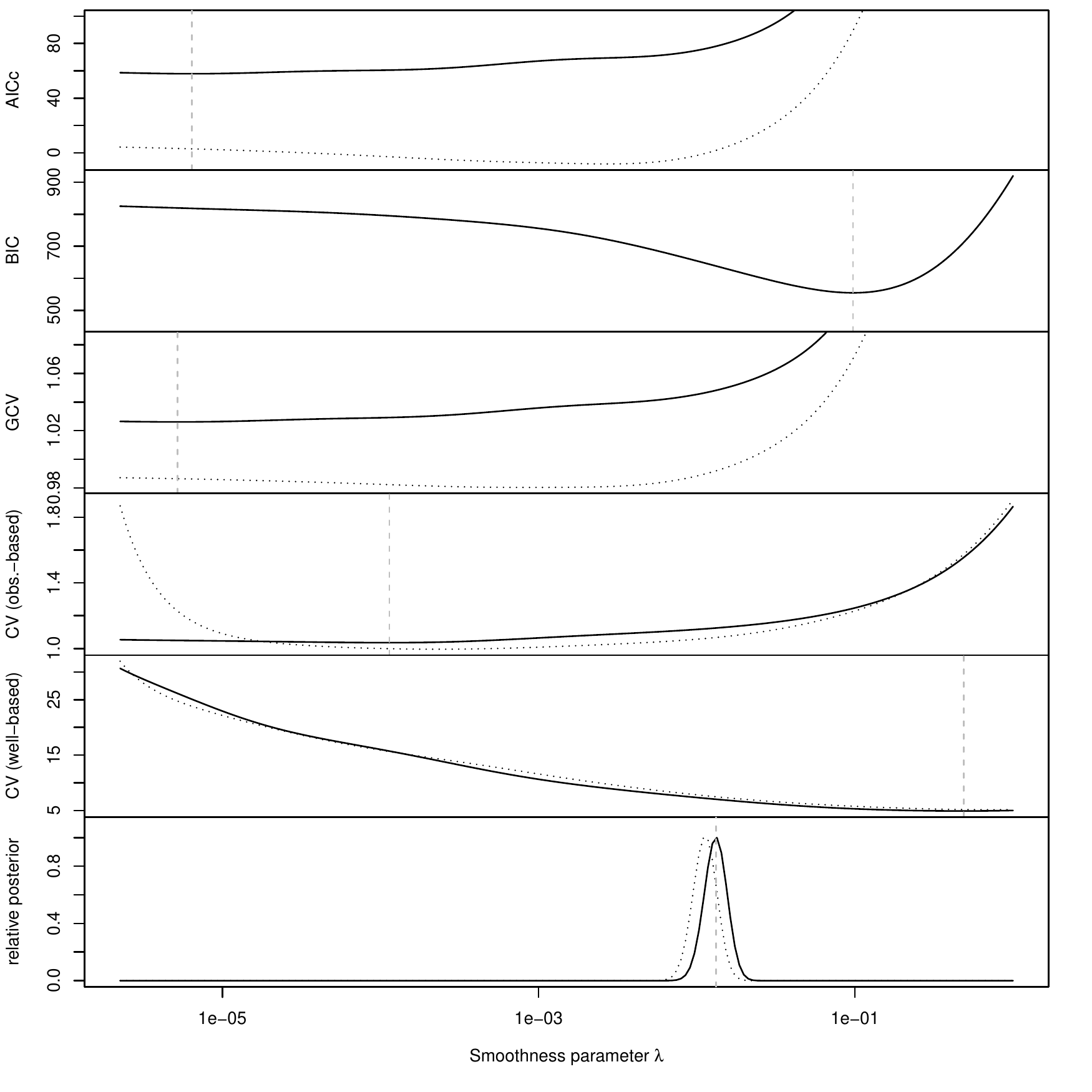}}
\caption{Different objective functions which can be used to determine the optimal amount of smoothing, applied to the Benzene data. The solid lines were obtained using all wells. The dotted lines were obtained after removing four wells.  The vertical dashed lines indicate the location of the minimum when all wells are used.}
\label{fig:criteria}
\end{figure}

Figure~\ref{fig:criteria} depicts the choice of the optimal value of the penalty parameter for this example.  In the case of the Bayesian {\sc map} approach, the optimal choice is given by the value of $\lambda$ which maximises its posterior distribution.  It is noticeable that {\sc aic}c, {\sc gcv} and observation-based cross-validation all lead to very low estimates of the smoothing parameter. These low values of the smoothing parameter effectively `switch off' the penalty, leading to `ballooning'. {\sc bic} and well-based cross-validation select a much larger value of the smoothing parameter, preventing `ballooning'. Using a fully Bayesian approach ({\sc MAP} or full model averaging) results in a smoothing parameter which is smaller than the one selected by {\sc bic} and well-based cross-validation, but which is still big enough to prevent `ballooning'. The posterior distribution of $\lambda$ is typically quite narrow, so using the {\sc map} usually gives results very similar to a fully Bayesian treatment of the smoothing parameter, which involves averaging over $\lambda$. This is borne out in this example, as well as in the simulations in section~\ref{sec:sim}.  As shown in figure \ref{fig:benzene} and discussed in section \ref{sec:psplines}, the {\sc map} is not sensitive to the removal of the four wells shown as crosses in the right-hand column of plots. In contrast, the values chosen by {\sc aic}c, {\sc gcv} and observation-based cross-validation are highly sensitive to the removal of the wells. Removing the wells results in a big change in the predictions, yielding much more credible predicted concentrations. This highlights that the well design plays a key role.

\subsection{Monitoring of {\sc mtbe} in groundwater}

\begin{figure}
\centerline{
   \includegraphics[width = 0.45\textwidth]{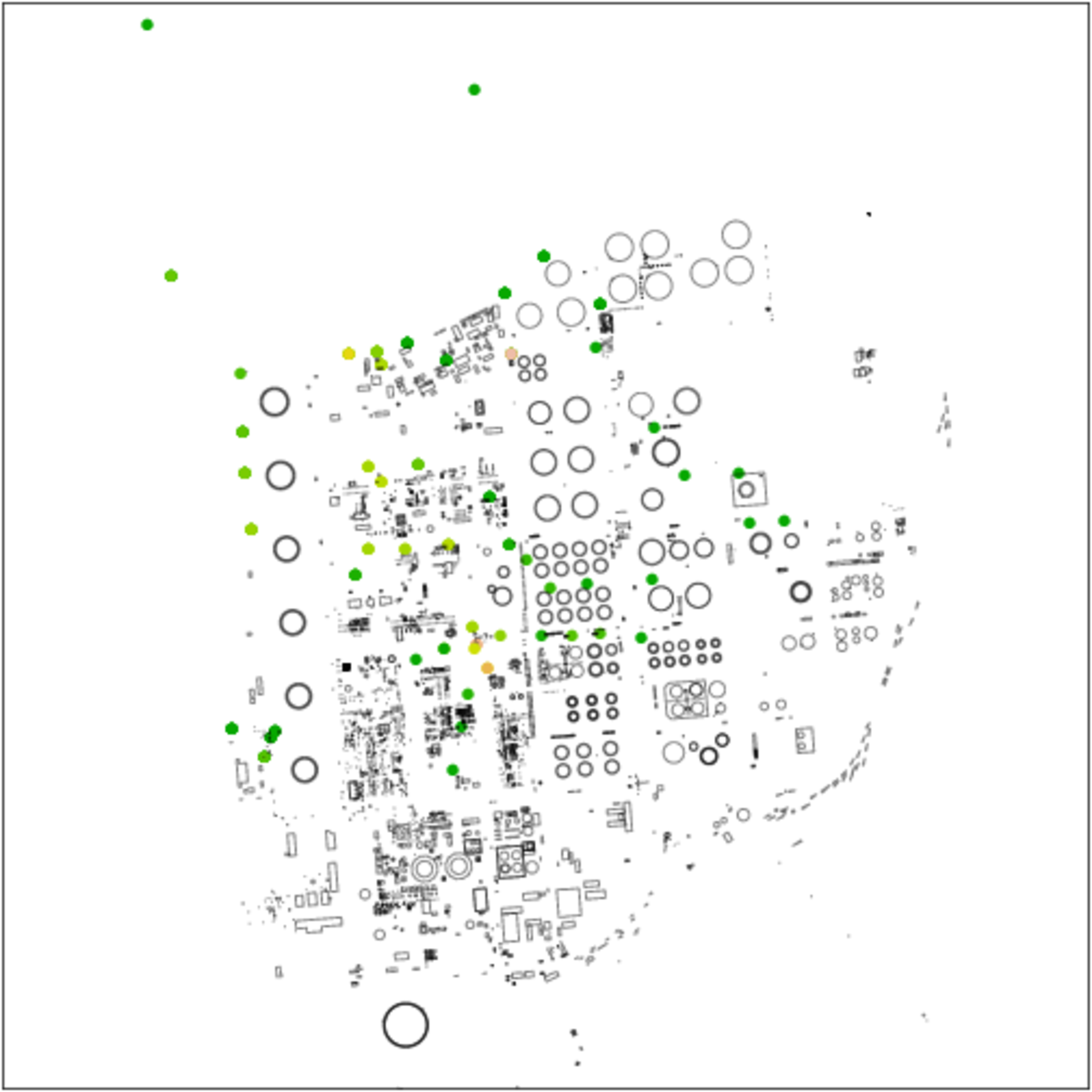}
    \includegraphics[width = 0.0715\textwidth]{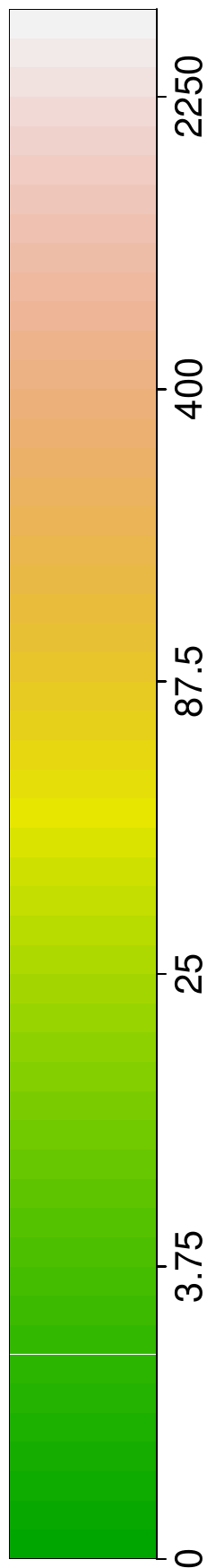}
}
\caption{Plan of the refinery site and wells. The wells are colour-coded according to observed concentrations of {\sc mtbe} immediately after release.}
\label{fig:mtbe0}
\end{figure}

\begin{figure}
   \subfigure[Predicted concentration surface for day 375]{\includegraphics[width = 0.45\textwidth]{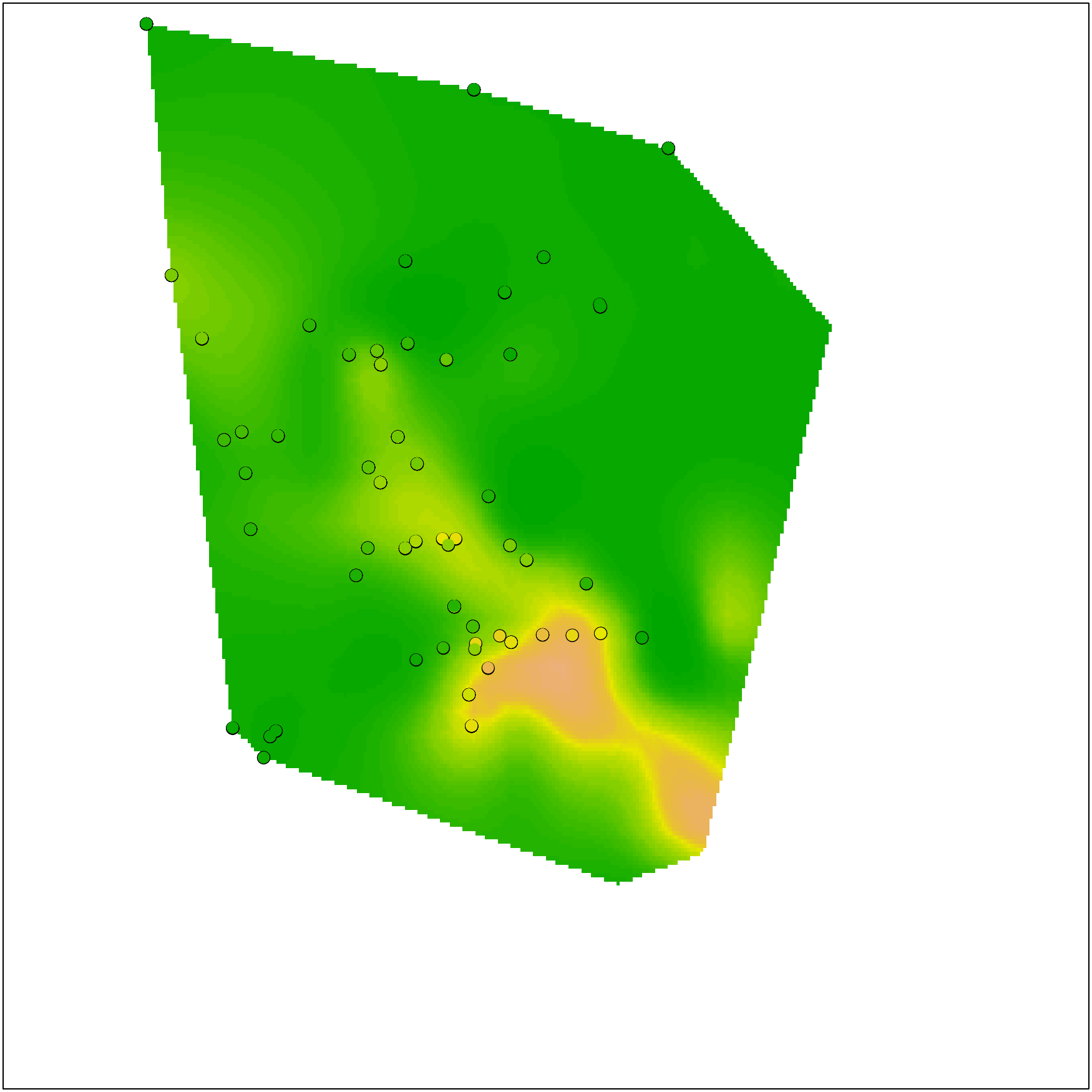}}
   \subfigure[Predicted concentration surface for day 700]{\includegraphics[width = 0.45\textwidth]{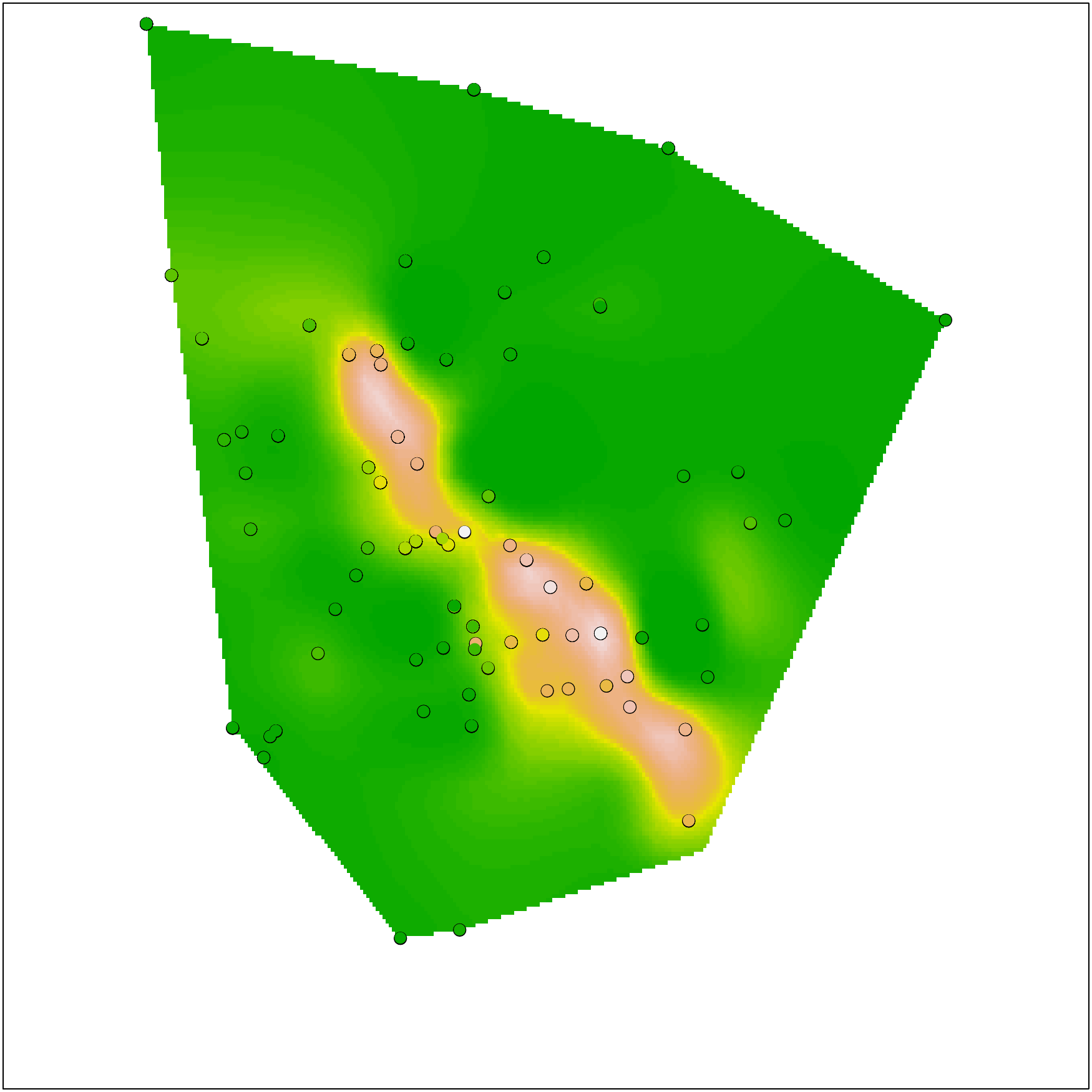}}
   
   \subfigure[Predicted concentration surface for day 900]{\includegraphics[width = 0.45\textwidth]{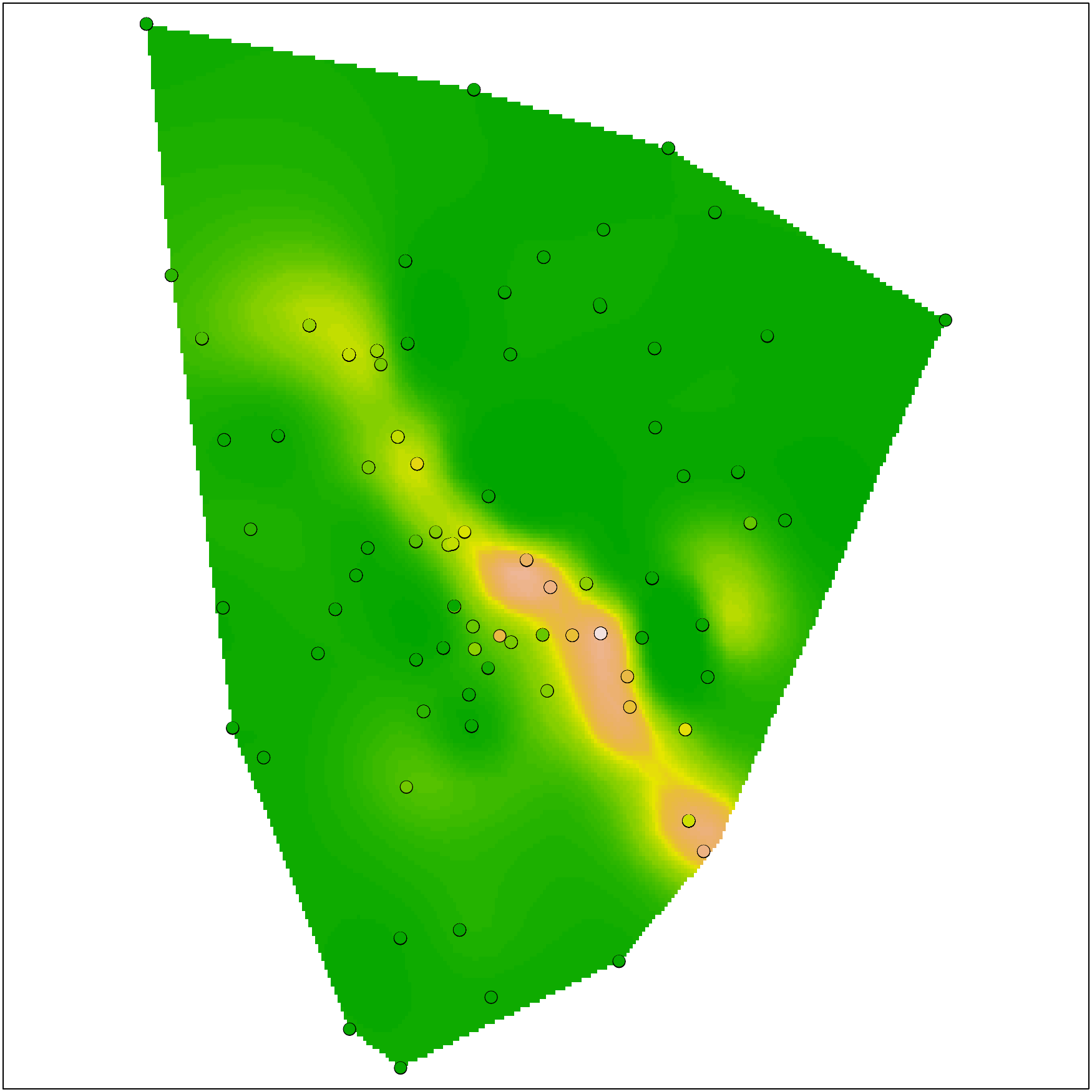}}
   \subfigure[Predicted concentration surface for day 1300]{\includegraphics[width = 0.45\textwidth]{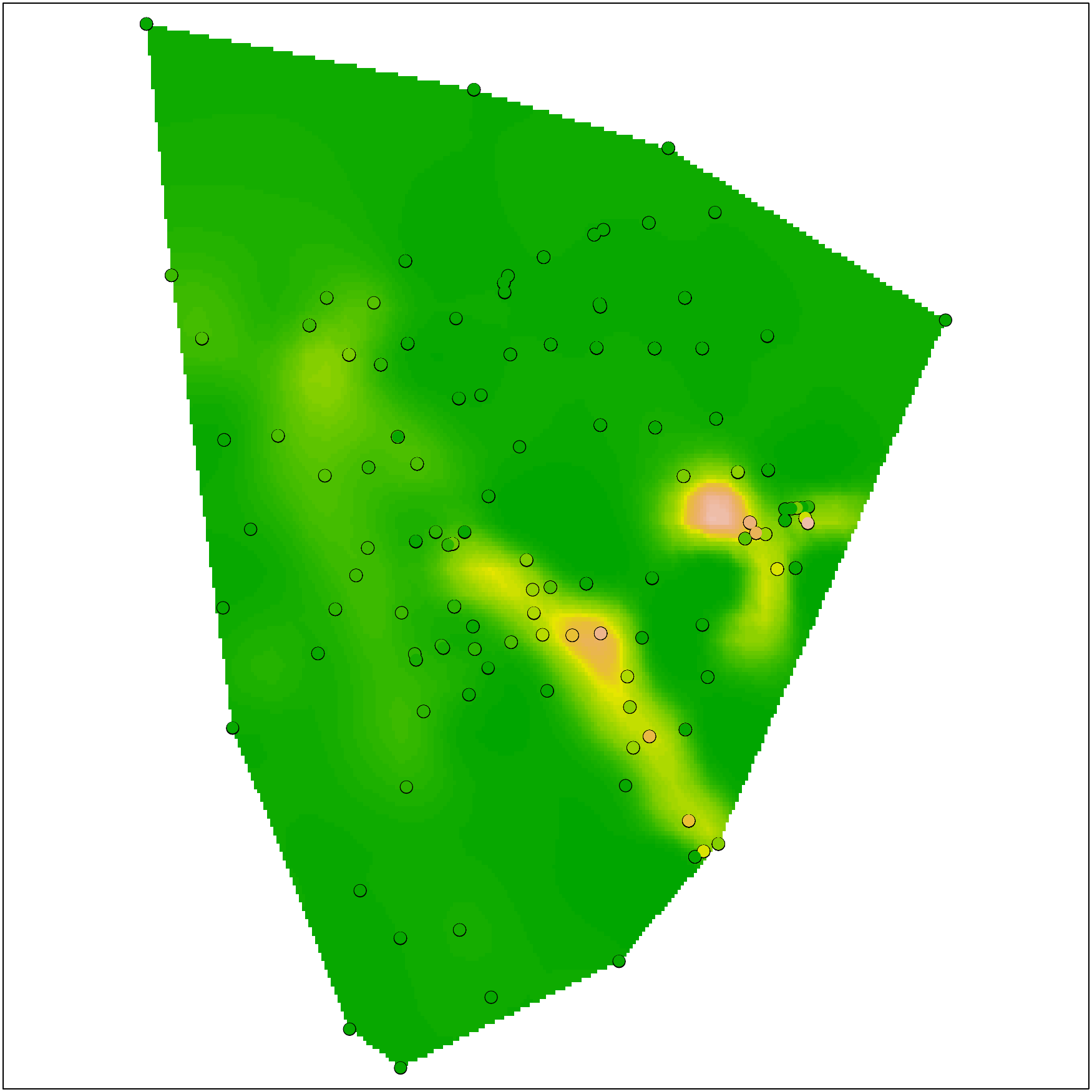}}
\caption{Predicted levels of {\sc mtbe} concentration obtained using the {\sc map} estimate of the smoothing parameter for four time points.  The colour scale is the same as that used in Figure~\ref{fig:mtbe0}}
\label{fig:mtbe}
\end{figure}

A more extensive example of the use of these techniques is provided by retrospective analysis of a dataset on a pollution event at a refinery site.  {\sc mtbe} (methyl tertiary butyl ether) is a petrol additive designed to reduce engine knocking and noxious emissions.  {\sc mtbe} is no longer in routine use at the site studied but was present in the refinery at the time of the event.  On entry to groundwater, {\sc mtbe} moves conservatively due to its high aqueous solubility and low retardation potential.  It degrades only slowly under anaerobic conditions.  Figure~\ref{fig:mtbe0} shows a schematic plan of the site with colour-coded points to indicate the concentrations of {\sc mtbe} measured at the monitoring wells at a date near the time of the {\sc mtbe} release.  Standard methods of analysis in this setting were to inspect individual well measurements over time to identify trends.  Geographical information systems were available and these were helpful for individual time snapshots but these could not easily be adapted to show the evolving dynamics of the incident.  Figure~\ref{fig:mtbe} (and the earlier Figures \ref{fig:benzene} and \ref{fig:sim}) was created using the \texttt{rp.spacetime} function from the current version of the \texttt{rpanel} \citep{Bowman:Crawford:Alexander:Bowman:2007:JSSOBK:v17i09} package for \texttt{R} \citep{R}.  This shows the estimated pollution surface using the Bayesian smoothing model described in Section~\ref{sec:bayesian}, using $18$ basis functions for easting, $22$ basis functions for northing, $14$ basis functions for time and the {\sc map} estimate of $\lambda$.  The shape and direction of the plume is clear and consistent with the south-east/north-west gradient in groundwater flow.  Despite the presence of protective pumping wells at the north-west boundary of the refinery site, the threat of {\sc mtbe} migrating across the site boundary and potentially reaching drinking water wells required immediate action.

The panels of Figure~\ref{fig:mtbe} show estimates from the spatiotemporal {\sc mtbe} distribution model at several further time points.  The first corresponds to the upgrading of a line of wells used to form a flow barrier in the middle of the site.  The effectiveness of these wells was greatly improved and the resulting curtailment of the plume to the north-west is apparent.  Subsequently, the source of the {\sc mtbe} release was identified near the south-east corner of the site and the model clearly tracks the dissipation and attenuation of {\sc mtbe} and the end of the incident.

\section{Discussion}
\label{sec:discussion}

A fully automatic Bayesian framework for determining the smoothing parameter in spatiotemporal p-spline models has been proposed. The focus was on a situation where the key objective was to deliver, on a fast timescale, automatic and robust estimates of the distribution of a solute in groundwater and the corresponding plume geometry. In particular, there was a need to avoid spurious local extrema of the predictive surface with little support in the data (`ballooning'), which can sometimes occur in regions where the well design is sparse. 

In our experience, and evidenced by the simulation study and the two real-world examples presented, the Bayesian methods studied are more stable than competing strategies based on crtieria such as the {\sc aic}c or {\sc gcv}. Whilst {\sc bic} is very good at avoiding `ballooning' it can also lead to over-smoothing, to which the other Bayesian methods are less prone. If used appropriately, which is difficult to judge without prior knowledge,  cross-validation can be very effective at preventing `ballooning', however it has a rather high computational cost.

Though our focus was on spatiotemporal models, the methods can also be applied to other smoothing problems in which the use of a single smoothing parameter is, possibly after rescaling of parts of the penalty or adjustment of the number of basis function, appropriate. 

The use of splines is not the only way of constructing spatiotemporal models for the contamination of groundwater. A particularly attractive alternative would be the use of a model based on the underlying physical processes.  However, such models require a good understanding of the geology of the site, which in turn requires additional information which is not always readily available.

In our experience the problem of `ballooning' is not limited to splines. `Ballooning' can also occur when using other techniques such as kriging with a Mat\'ern covariance. In the latter case the severity of the problem of `ballooning' depends on how the shape parameter of the covariance function is chosen.

The methodology set out above is implemented in {\sc gwsdat}, a fully automatic tool for the analysis of groundwater contaminants developed by Shell Global Solutions.

\section{Acknowledgements}

This work was partly funded by Shell Global Solutions (UK) Ltd. The views expressed are those of the authors and may not reflect the policy or position of their employing organisations.

\bibliographystyle{apalike}
\bibliography{MAP-paper}

\appendix

\section{Computational details}

Without loss of generality we shall assume that $\Db$ has full row rank. If not, replace $\Db$ by the non-zero rows of $\Rb$ from a QR decomposition $\Db=\Qb\Rb$.

If the differencing matrix $\Db$ is of full column rank we could reparametrise the problem using $\tilde\alphab=\Db\alphab$ yielding the ridge-regression problem
$$
\yb| \tilde\alphab \sim \mathcal{N}(\Bb\Db^{-1}\tilde\alphab,\sigma^2\Ib),\qquad  \tilde\alphab\sim \mathcal{N}(\zerob,\sigma^2\lambda^{-1}\Ib)
$$
which can be solved independently of $\lambda$ using a singular value decomposition \citep[see e.g.][]{golub-vanloan:1996}.

However, in P-spline problems the matrix $\Db$ does not have full column rank. Thus we have to identify the subspace of the covariate space on which the penalty does not act. This can be done using a QR decomposition of $\Db'=(\tilde\Qb_1,\tilde\Qb_2)\left(\begin{array}{r}\tilde\Rb_1\\\zerob\end{array}\right)$. Setting
$\tilde \alphab_1=\tilde\Rb_1'\tilde\Qb_1'\alphab$, $\tilde \alphab_2=\tilde\Qb_2'\alphab$, $\tilde\Bb_1=\Bb\tilde\Qb_1\tilde{\Rb_1'}{}^{-1}$, $\tilde\Bb_2=\Bb\tilde\Qb_2$ and $\tilde\Db_1=\tilde\Rb'$ allows rewriting the problem as
$$
\yb|\tilde\alphab_1,\tilde\alphab_2 \sim \mathcal{N}(\tilde\Bb_1\tilde\alphab_1 + \tilde\Bb_2\tilde\alphab_2 ,\sigma^2\Ib),\qquad  \tilde\alphab_1\sim \mathcal{N}(\zerob,\sigma^2\lambda^{-1}\Ib),\qquad p(\tilde\alphab_2)\propto 1.
$$
This rotation of the variable space allows the regression coefficient to be split into two vectors: one, $\tilde \alphab_1$, with a proper standard normal prior and one, $\tilde\alphab_2$, with an improper flat prior.

Finally, we perform a rotation of the observation space such that the design matrix of $\tilde\alphab_2$ only has $l$ non-zero rows, where $l$ is the length of the vector $\tilde \alpha_2$. This is achieved by computing the QR decomposition of $\tilde\Bb_2=(\breve \Qb_1,\breve \Qb_2)\left(\begin{array}{r}\breve \Rb_1\\\zerob\end{array}\right)$.

Considering the response $\breve \yb=(\breve \yb_1',\breve\yb_2')'=(\breve \Qb_1,\breve \Qb_2)'\yb$ as well as setting
$\breve \Bb_{11}=\breve \Qb_1\tilde\Bb_1$, $\breve \Bb_{21}=\breve \Qb_2\tilde\Bb_1$ and
$\breve \Bb_{12}=\breve \Qb_1\tilde\Bb_2$ yields the equivalent model assumption

$$
\left.\left(\begin{array}{r}\breve\yb_1\\\breve\yb_2\end{array}\right)
\right|\tilde\alphab_1,\tilde\alphab_2
\sim \mathcal{N}\left(
\left(\begin{array}{cc}
\breve\Bb_{11}&\breve \Bb_{12}\\
\breve\Bb_{21}&\zerob
\end{array}\right)
\left(\begin{array}{r}\tilde\alphab_1\\\tilde\alphab_2\end{array}\right)
,\sigma^2\Ib\right)
$$
with the same priors for $\tilde\alphab_1$ and $\tilde\alphab_2$ as above. The posterior distribution of $\tilde\alphab_1$ and $\tilde\alphab_2$ given $\breve \yb$ is a normal distribution with the penalised least-squares estimates $\widehat{\tilde\alphab_1}$ and $\widehat{\tilde\alphab_2}$ as mean and unscaled covariance matrix
$$
\left(\begin{array}{cc}
\breve\Bb_{11}'\breve\Bb_{11}+\breve\Bb_{21}'\breve\Bb_{21}+\lambda \Ib & \breve\Bb_{11}'\breve\Bb_{12}\\
\breve\Bb_{12}'\breve\Bb_{11}& \breve\Bb_{12}'\breve\Bb_{12}
\end{array}\right).
$$
Its determinant can be shown to be
$$
\det(\breve\Bb_{21}'\breve\Bb_{21}+\lambda \Ib)\cdot\det(\breve\Bb_{12}'\breve\Bb_{12})
$$
by using the Leibnitz formula and by exploiting the fact that by construction $\breve\Bb_{12}$ is invertible. The penalised least squares estimate can be found by exploiting the fact that the residual corresponding to $\breve\yb_1$ can be set to $\zerob$ by setting
$$
\widehat{\alphab_2}=\breve\Bb_{12}^{-1}\left(\breve\yb_2-\breve\Bb_{11}\widehat{\tilde\alphab_1}\right).
$$
Thus $\widehat{\tilde\alphab_1}$ can be found by considering the reduced ridge regression problem involving only only the coefficient $\tilde \alphab_1$:
$$
\breve\yb_2|\tilde\alphab_1 \sim \mathcal{N}(\breve\Bb_{21}\tilde\alphab_1,\sigma^2\Ib),\qquad \tilde\alphab_1\sim \mathcal{N}(\zerob,\sigma^2\lambda^{-1}\Ib)
$$
This is a ridge regression problem which can be solved using the singular value  decomposition of $\breve\Bb_{21}=\Ub\Lb\Vb'$ yielding \citep[see e.g.][]{golub-vanloan:1996}
$$\widehat{\tilde\alphab_1}=\Vb\left(\frac{\textrm{diag}(\Lb)\odot\Ub'\breve\yb_2}{\textrm{diag}(\Lb)^2 +\lambda}\right)
$$
where $\odot$ stands for element-wise multiplication and the fraction, sum and power are to be interpreted element-wise. The singular value decomposition can also be used to compute the determinant
$$
\det(\breve\Bb_{21}'\breve\Bb_{21}+\lambda \Ib)=\det(\Lb'\Lb+\lambda \Ib)=
\prod_i (L_{ii}+\lambda)
$$

Up to the computation of the final singular value decomposition all operations can be performed by exploiting the sparsity of the matrices.  It is also worth noting that  it is not necessary to compute a full SVD. It is enough to tridiagonalise the matrix $\breve\Bb_{21}$ \citep[see e.g.][]{Elden-1977,wood-2000-jrssb}. Exploiting this allows to further speed-up the algorithm.

\end{document}